\documentstyle[12pt]{article}
\begin{document}

\title{\bf\huge
Algebraic Bethe ansatz for eight vertex model with general 
open-boundary conditions
}

\author{
{\bf
Heng Fan$^{a,b}$,Bo-yu Hou$^b$,Kang-jie Shi$^{a,b}$,Zhong-xia Yang$^b$}\\
\normalsize
$^a$ CCAST(World Laboratory)\\
P.O.Box 8730,Beijing 100080,China\\
$^b$ Institute of Modern Physics,Northwest University\\
P.O.Box 105,Xian,710069,China
}

\maketitle

\begin{abstract}
By using the intertwiner and face-vertex correspondence relation, we
obtain the Bethe ansatz equation of eight vertex model with open
boundary conditions in the framework of algebraic Bethe ansatz method.
The open boundary condition under consideration is 
the general solution of the reflection equation
for eight vertex model with only one resctriction on 
the free parameters of
the right side reflecting boundary matrix. The reflecting 
boundary matrices used in this paper thus may have off-diagonal elements.
Our construction can also be used for the Bethe ansatz of SOS model with 
reflection boundaries.
\end{abstract}
\vskip 2truecm
PACS: 7510J, 0520, 0530.


\newpage

\section{Introduction}
One of the most important goal of exactly solvable lattice models is
to find the eigenvalues and eigenvectors of the transfer
matrix of a system, then to obtain the
thermodynamic limit of this system.

Bethe examined the completely isotropic case of the $XXX$ model and found
the eigenvalues and eigenvectors of its Hamiltonian[1]. After Bethe's work,
Yang and Yang analyzed the anisotropic $XXZ$ 
model by means of Bethe ansatz[2].
Then, Baxter in his remarkable papers gave a solution for the completely
anisotropic $XYZ$ model[3]. He discovered a relation between the quantum
$XYZ$ model and eight vertex model which is 
one of the two dimensional exactly solvable lattice model.
Faddeev and Takhtajan simplified Baxter's formulae and
proposed the quantum inverse scattering method or algebraic
Bethe ansatz method to solve the six vertex and eight vertex models, whose
spin chain equivalent are $XXZ$ spin model and $XYZ$ spin model, 
respectively[4]. After Yang-Baxter-Faddeev-Takhtajan's work, a lot of exactly
solvable models have been solved by algebraic Bethe ansatz[5,6], functional
Bethe ansatz[7,8], co-ordinate Bethe ansatz[9], etc.
                                     
Typically, the two-dimensional exactly solvable lattice models are solved
by imposing periodic boundary conditions in which the Yang-Baxter equation
provides a sufficient condition for the integrability of the models. 
\begin{eqnarray}
R_{12}(z_1-z_2)R_{13}(z_1-z_3)R_{23}(z_2-z_3)
=R_{23}(z_2-z_3)R_{13}(z_1-z_3)R_{12}(z_1-z_2)
\end{eqnarray}
\noindent where, the R-matrix is the Boltzmann weight for the vertex models
in two dimensional statistical mechanics. As usual, $R_{12}(z),R_{13}(z)$
and $R_{23}(z)$ act in $C^n\otimes C^n\otimes C^n$ with $R_{12}(z)
=R(z)\otimes 1, R_{23}(z)=1\otimes R(z),$etc.

The exactly solvable models with non-periodic boundary conditions 
have been early studied in ref.[10-14].
Recently, integrable models with open boundary conditions have been 
attracting a great deal of interests. This was initiated by Cherednik[15]
and
Sklyanin[16], they proposed a systematic approach to handle the open
boundary condition problems which involves the so called reflection 
equation (RE).
\begin{eqnarray}
& &R_{12}(z_1-z_2)K_1(z_1)R_{21}(z_1+z_2)K_2(z_2)\nonumber \\
&=&K_2(z_2)R_{12}(z_1+z_2)K_1(z_1)R_{21}(z_1-z_2)
\end{eqnarray}
\noindent The open boundary conditions are determined by the boundary 
reflecting matrix $K$ satisfying the RE.

By using a non-trivial generalization of the quantum inverse scattering
method, Sklyanin obtained the Bethe ansatz equation of six vertex model
with open boundary conditions by algebraic Bethe ansatz method[16]. The 
transfer matrix with a particular choice of boundary conditions is
quantum group $U_q(sl(2))$ invariant[17,18]. After Sklyanin's pioneering
work, a lot of exactly solvable lattice models with open boundary
conditions have been solved. Mezincescu and Nepomechie solved the
$A_1^{(1)}$ and $A_2^{(2)}$ vertex models by using the fusion procedure[19].
Foerster and Karowski solved $spl_q(2,1)$ invariant Hamiltonian which
contains a non-trivial boundary term by 
using the nested algebraic Bethe ansatz[20].
Using the same method, de Vega and Gonzalez-Ruiz solved the $A_n$ vertex
model, they also analyzed the thermodynamic limit of this system[21]. 
Yue, Fan
and Hou solved the general $SU_q(n|m)$ vertex model[22]. For other progress
about open boundary conditions along this direction, see e.g. ref.[23-31].

However, for Baxter's eight vertex model with open-boundary conditions, 
little progress has been made. Jimbo et al obtained the difference
equation of n point function for semi-infinite $XYZ$ chain [32].
Yu-kui Zhou have recently studied the fused eight 
vertex model and have found the
functional relations for eight vertex model with open 
boundary conditions[33],
in which the reflecting $K$ matrix which satisfy the vertex
RE(2) are diagonal.

In this paper, we will use the algebraic Bethe ansatz method to solve the
eight vertex model with open boundary conditions. It is known that $K$ 
matrix is a solution of RE. The general solution of RE for six vertex 
model was obtained by de Vega[34]. In ref.[35,36], solutions of RE for eight
vertex model have been found. The general solution of RE for eight vertex
model which has three free parameters was found by two groups[37,38].
In our approach we use the general solution for left and right boundaries.
The reflecting boundary $K$ matrices thus may have off diagonal elements. We
need only to impose one relation on the free parameters of 
the right $K$ matrix, if the free parameters of the left $K$ matrix 
are arbitrarily given.

It is known that in Baxter's original work, in stead of Yang-Baxter relation,
the star-triangular relation plays the key role which can be obtained
from Yang-Baxter relation by using the intertwiners. This 
was later generalized to $Z_n$ Baxter-Belavin model[39,40] to 
describe the
interaction-round-a-face model by 
Jimbo, Miwa and Okado[41]. This intertwiner method was also used
to solve the Bethe ansatz problem for similar cases [42,43].
In order to get algebraic Bethe ansatz equation of eight vertex model with
open-boundary conditions, we need to describe the exchange relations of 
the monodromy matrix in the "face language". Thus in our paper, we need
to convert our boundary conditions of vertex model to 
that of a face model. Our approach
is equivalent to a SOS model with open boundaries satisfying the face RE.
The face RE was first proposed
by Behrend, Pearce and Brien [44]. In ref.[44], they also find a 
diangonal solution of face RE for
ABF[45] model. By using the intertwiners, the face RE is derived directly
from vertex RE and the general solution of face RE for eight vertex
SOS model are found by other groups[46,33]. In this paper we actually
use a diagonal solution of face RE at the left boundary and 
an upper triangular solution at the right boundary for
eight vertex SOS model. 
This open boundary conditions are different from the case 
discussed by Yu-kui Zhou in ref.[33]. 
Since a diagonal matrix is a special
case of the upper triangular matrix, our approach can be used to a 
SOS model with boundaries proposed by Behrend et al. 
We can prove that by taking a special case, the
Bethe ansatz equation for RSOS model, ABF model, can also be obtained.

The outline of this paper is as follows. In section 2, we first review the 
eight vertex model and reflecting open boundary conditions, the model
under consideration in this paper will also be constructed. In sect.3, by
using the correspondence of face and vertex models, we derive the face RE. 
As mentioned above, the face RE will play a key role in the algebraic
Bethe ansatz method for eight vertex model instead of vertex RE. The
transfer matrix with boundary conditions will also be constructed by
using the face weight. In sect.4, we will find the local vaccum for eight
vertex model with a boundary which is the same as the vaccum state given
by Baxter[3,7] in the Bethe ansatz of eight vertex model with periodic
boundary condition. The face boundary matrix derived directly from vertex
boundary matrix is obtained. In sect.5, the Bethe ansatz problem is solved
for eight vertex model with open-boundary conditions. Section 6 contains
some discussions and the further work.

\section{Description of the model}
\subsection{The R matrix}

We first start from the R-matrix of the eight vertex model. Denote
$\alpha =(\alpha _1,\alpha _2), \alpha _1,\alpha _2=0,1$. Let $g$ and
$h$ be $2\times 2$ matrices with elements $g_{ii'}=(-1)^i\delta _{ii'},
h_{ii'}=\delta _{i+1,i'}, i,i'=0,1$. 

Define $2\times 2$ matrices $I_{\alpha }=I_{(\alpha _1,\alpha _2)}
=h^{\alpha _1}g^{\alpha _2}, I_0=I=identity$, and define 
$I_{\alpha }^{(j)}
=I\otimes I\otimes \cdots I_{\alpha }\otimes \cdots \otimes I,I_{\alpha }$
is at $j$-th space. As usual, $I_{\alpha }^{(j)}$ act in $V=V_1
\otimes \cdots \otimes V_l$, where space $V$ is consisted of $l$ 
two-dimensional spaces. We then introduce some notations used 
in this paper.They are
\begin{eqnarray}
\theta \left[ \begin{array}{c}
               a\\b \end{array} \right]
(z,\tau )&\equiv &\sum _{m\in Z}exp \left\{ \pi 
{\sqrt {-1}}(m+a)[(m+a)\tau +2(z+b)]
\right\},\\
\sigma _{\alpha }(z)&\equiv &\theta \left[ \begin{array}{c}
\frac {1}{2}+\frac {\alpha _1}{2}\\
\frac {1}{2}+\frac {\alpha _2}{2}\end{array}\right] (z,\tau),\\
h(z)&\equiv &\sigma _{(0,0)}(z),\\
\theta ^{(i)}(z)&\equiv &\theta \left[ \begin{array}{c}
{1\over 2}-{i\over  2}\\ {1\over 2} \end{array}\right ] (z,2\tau ),i=0,1,
\\
W_{\alpha }(z)&=&{1\over 2}\frac {\sigma _{\alpha }(z+{w\over 2})}
{\sigma _{\alpha }({w\over 2})}.
\end{eqnarray}
\noindent The R-matrix of eight vertex model takes the form (Fig.1a)
\begin{eqnarray}
R_{jk}(z)=\sum _{\alpha }W_{\alpha }(z)I_{\alpha }^{(j)}
(I_{\alpha }^{-1})^{(k)}
\end{eqnarray}
\noindent which satisfy the Yang-Baxter equation (Fig.2a)
\begin{eqnarray}
& &R_{ij}(z_i-z_j)R_{ik}(z_i-z_k)R_{jk}(z_j-z_k)\nonumber \\
&=&R_{jk}(z_j-z_k)R_{ik}(z_i-z_k)R_{ij}(z_i-z_j).
\end{eqnarray}

\noindent It can be proved that the R 
matrix of eight vertex model satisfy the following 
unitarity and cross-unitarity conditions,

\begin{eqnarray}
{\rm unitarity}&:& R_{ij}(z)R_{ji}(-z)=\rho (z)\cdot id,\\
{\rm cross-unitarity}&:& R_{ij}^{t_i}(z)R_{ji}^{t_i}(-z-2w)=\rho '(z)
\cdot id,
\end{eqnarray}
\noindent where $id$ is the identity and 
$t_i$ denotes transposition in the $i$-th space. $\rho (z)$
and $\rho '(z)$ are scalars satisfying 
\begin{eqnarray}
\rho (z)=\rho (-z),\\
\rho '(z)=\rho '(-z-2w).
\end{eqnarray}
\noindent In eqs.(8-11), the indices take the value $i,j,k=1,\cdots ,l$.
 
\subsection{Reflection equation and reflecting boundary conditions}
We now deal with an exactly solvable lattice model with reflecting boundary
conditions, the R-matrix defined above is the Boltzmann weights for this
lattice model. In order to construct the transfer matrix of this system, we
must introduce a reflecting boundary matrix $K(z)$ which is a $2\times 2$
matrix and satisfy the reflection equation proposed by Cherednik[15]
and Sklyanin[16] (Fig.2b).

\begin{eqnarray}
& &R_{12}(z_1-z_2)K_1(z_1)R_{21}(z_1+z_2)K_2(z_2)\nonumber \\
&=&K_2(z_2)R_{12}(z_1+z_2)K_1(z_1)R_{21}(z_1-z_2)
\end{eqnarray}

\noindent As mentioned in the introduction, two groups have obtained 
independently the general solution $K$ of this reflection equation
for eight vertex model. Here
we take the solution $K(z)$ as [38]
\begin{eqnarray}
K(z)=\sum _{\alpha }C_{\alpha }\frac {I_{\alpha }}{\sigma _{\alpha }(-z)},
\end{eqnarray}
\noindent where $C_{\alpha }$ are arbitrary parameters. Correspondingly,
we have the dual reflection equation which is necessary in the following
of this paper (Fig.2c): 

\begin{eqnarray}
& &R_{12}(z_2-z_1){\tilde {K}}_1(z_1)R_{21}(-z_1-z_2-2w)
{\tilde {K}}_2(z_2)\nonumber \\
&=&{\tilde {K}}_2(z_2)R_{12}(-z_1-z_2-2w){\tilde {K}}_1(z_1)
R_{21}(z_2-z_1).
\end{eqnarray}
 
\noindent We take the solution of this reflection equation as 
\begin{eqnarray}
\tilde {K}(z)=\sum _{\alpha }\tilde {C}_{\alpha }\frac {I_{\alpha }}
{\sigma _{\alpha }(z+w)},
\end{eqnarray}
\noindent where $\tilde {C}_{\alpha }$ are also arbitrary parameters.
Usually, we also call the reflecting boundary $K$ and $\tilde {K}$ matrices
as right and left boundary matrices, respectively. 

In order to deal with the systems with open boundary conditions, let us
define two forms of standard "row-to-row" monodromy matrices $S_1(z_1)$
and $T_1(z_1)$ which act in the space $V=V_1\otimes V_2\otimes \cdots
\otimes V_l$ by

\begin{eqnarray}
S_1(z_1)&=&R_{l1}(u_l+z_1)R_{l-1,1}(u_{l-1}+z_1)\cdots 
R_{31}(u_3+z_1),\nonumber \\
T_1(z_1)&=&R_{13}(z_1-u_3)R_{14}(z_1-u_4)\cdots R_{1l}(z_1-u_l),
\end{eqnarray}
\noindent where $(u_l,u_{l-1},\cdots u_3)\equiv \left\{ u_i\right\} $
are arbitrary parameters. $S_2(z_2)$ and $T_2(z_2)$ can also
be similarly defined.

Considering $\left\{ u_i\right\} $ are the same for $S_1(z_1)$ and $S_2(z_2)$,
and noticing that two R matrices acting on four different spaces
commute with each other, we find
\begin{eqnarray}
R_{21}(z_1-z_2)S_1(z_1)S_2(z_2)=R_{21}(z_1-z_2)R_{l1}(u_l+z_1)
R_{l2}(u_l+z_2)\cdots . 
\end{eqnarray}
\noindent Using Yang-Baxter equation repeatedly, we have
\begin{eqnarray}
R_{21}(z_1-z_2)S_1(z_1)S_2(z_2)=S_2(z_2)S_1(z_1)R_{21}(z_1-z_2).
\end{eqnarray}
\noindent Similarly, we can also obtain
\begin{eqnarray}
T_1(z_1)R_{12}(z_1+z_2)S_2(z_2)=S_2(z_2)R_{12}(z_1+z_2)T_1(z_1),\\
T_2(z_2)T_1(z_1)R_{12}(z_1-z_2)=R_{12}(z_1-z_2)T_1(z_1)T_2(z_2).
\end{eqnarray}

For the periodic boundary condition cases which are studied extensively 
before, the transfer matrix is defined as the trace of the standard
"row-to-row" monodromy matrix. But for the open boundary conditions cases,
instead of the standard "row-to-row" monodromy matrix, we should define
the "double-row" monodromy matrices which take the form:

\begin{eqnarray}
k_1(z_1)&=&T_1(z_1)K_1(z_1)S_1(z_1),\nonumber \\
k_2(z_2)&=&T_2(z_2)K_2(z_2)S_2(z_2).
\end{eqnarray}

\noindent Using the relations listed above, we can prove $k_i(z_i)$ satisfy
the reflection equation
\begin{eqnarray}
& &R_{12}(z_1-z_2)k_1(z_1)R_{21}(z_1+z_2)k_2(z_2)\nonumber \\
&=&k_2(z_2)R_{12}(z_1+z_2)k_1(z_1)R_{21}(z_1-z_2).
\end{eqnarray}

\noindent As used usually in the framework of the quantum inverse
scattering method, $k_i(z_i)$ are $2\times 2$ matrix with elements defined
as operators acting in the space $V'=V_3\otimes V_4\otimes \cdots \otimes V_l$
which is the so called quantum space, the spaces $V_1$ and $V_2$ are
the auxiliary spaces. Eqs.(14) and (24) 
shows that $k(z)$ is the co-module of 
$K(z)$.
 
\subsection{The transfer matrix}
Now, let's formulate the transfer matrix with open boundary conditions.
\begin{eqnarray}
t(z_i)&=&Tr_{V_i}\left\{ {\tilde {K}}_i(z_i)k_i(z_i)\right\} \nonumber \\
&=&\sum _{kl} {\tilde {K}}(z_i)_{kl}k(z_i)_{lk},
\end{eqnarray}
\noindent with $i=1,2$. Since the transfer 
matrices are defined as the trace over
the auxiliary spaces $V_i, i=1,2$, they should be independent of $V_1$ 
and $V_2$, and
are represented as operators acting in the quantum space $V_3\otimes \cdots
\otimes V_l$. With the help of the unitarity, cross-unitarity relations of R
matrix, Yang-Baxter relation, reflection equation and its dual reflection
equation, we can prove that the transfer matrices with different spectrum 
commute with each other[47],
\begin{eqnarray}
t(z_1)t(z_2)=t(z_2)t(z_1).
\end{eqnarray}
\noindent This ensures the integrability of this system.

The aim of this paper is to find the eigenvalues and eigenvectors of the 
transfer matrix which defines the Hamiltonian of the system under 
consideration. We will use the algebraic Bethe ansatz method to solve this
problem. The transfer matrix is defined as a linear function of 
the elements of the "double-row" monodromy matrix. 
So, it is necessary to find the proper linear combinations of
the elements of the "double-row" monodromy matrix whose commutation
relations are suitable for algebraic Bethe ansatz. 
Besides this, we also need to find an "vaccum"
state which is independent of the spectrum $z$. It is well known that
this "vaccum" state can be obtained easily for six vertex model with
periodic- or open-boundary conditions. For eight vertex model, it is
not a trivial problem. We will study the commutation relations and
the "vaccum" state problems in the following sections.

\section{Commutation relations}
It is known that for six vertex model and other
trigonometric vertex models we can obtain the necessary commutation 
relations directly from
the reflection equation in which $k(z)$ is the "double-row"
monodromy matrix. 
But for the eight vertex model whose R matrix has eight non-zero
elements, we can not obtain such relations directly from the reflection 
equation. We have to use the vertex-face correspondence to solve this
problem.  That means we should properly combine the elements of $k(z)$ 
so that we can find  
simple commutation relations  
which can be dealed with by
algebraic Bethe ansatz method.
\subsection{Face-vertex correspondence}
We first define a two element column vectors $\phi _{m ,\mu }(z),
\mu =0,1, m\in Z$, whose $k$-th element is [7,41,49] (Fig.1c) 
\begin{eqnarray}
\phi _{m,\mu }^k
=\theta ^{(k)}(z+(-1)^{\mu }wa+w\beta ),
\end{eqnarray}
\noindent where $a=m+\gamma ,\gamma ,\beta \in C, k=0,1$. We call $m$ the
face weight,$\mu $ the face index, which take
values $0, 1$. $\phi $ is usually called
the three-spin operator. It can be proved that we can find row vectors
$\bar {\phi }, \tilde {\phi }$ satisfying the following conditions for
generic $w, \beta, \gamma $.

\begin{eqnarray}
\tilde {\phi }_{m+\hat {\mu },\mu }(z)\phi _{m+\hat {\nu },\nu }(z)
=\delta _{\mu \nu },\nonumber \\
\bar {\phi }_{m,\mu }(z)\phi _{m,\nu }(z)=\delta _{\mu \nu },
\end{eqnarray}
\noindent where 
\begin{eqnarray}
\hat {\mu }\equiv (-1)^{\mu },\hat {\nu }\equiv
(-1)^{\nu }.\nonumber  
\end{eqnarray}
\noindent The above relations can also be written in other forms
\begin{eqnarray}
\sum _{\nu =0}^1\phi _{m+\hat {\nu },\nu }(z)\tilde {\phi }_{m+
\hat {\nu },\nu }(z)=I,\nonumber \\
\sum _{\mu =0}^1\phi _{m,\mu }(z)\bar {\phi }_{m,\mu }(z)=I.
\end{eqnarray}
\noindent As usual, $I$ is the $2\times 2$ unit matrix. (Fig.3)

We define the face Boltzmann weights for the 
interaction-round-a-face model (IRF) as follows[7,41,45,49]:
\begin{eqnarray}
W(m|z)^{\mu \mu }_{\mu \mu }&=&\frac {h(z+w)}{h(w)}\nonumber \\
W(m|z)^{\nu \mu }_{\mu \nu }&=&\frac {h( w(m+\gamma )-(-1)^{\mu }z) }
{h(w(m+\gamma ))},\mu \not= \nu ,\nonumber \\
W(m|z)^{\mu \nu }_{\mu \nu }&=&\frac {h(z)h(w(m+\gamma )-(-1)^{\mu }w)}
{h(w)h(w(m+\gamma ))}, \mu \not= \nu ,
\end{eqnarray}
\noindent where the face indices $\mu ,\nu $ take the values $0,1$.
The other face Boltzmann weights are defined as zeroes, so we can see 
explicitly that for a given face weight $m$, 
we only have six non-zero face Boltzmann weights at all. 
Traditionally, the face Boltzmann weights 
for eight vertex SOS model are denoted
as $W_z\left[ \begin{array}{cc}
a & b\\
c &d \end{array} \right] $. Its relation with the notations
used in this paper is (Fig.1b):
\begin{eqnarray}
W(m|z)^{\mu '\nu '}_{\mu \nu }=W_z\left[ \begin{array}{cc}
m+\hat {\mu }+\hat {\nu } & m+\hat {\nu }'\\
m+\hat {\mu } & m \end{array} \right] 
\end{eqnarray}

\noindent The reasons that we use notations (30) are that 
a lot of zero face Boltzmann 
weights will not appear in our calculation. It is also convenient to
compare the results of eight vertex model case with the six vertex model
case.

The face Boltzmann weights of IRF model 
defined above have a relation with the R-matrix
of eight vertex model which is usually called the face-vertex 
correspondence[3,7,41].

\begin{eqnarray}
& &R_{12}(z_1-z_2)\phi _{m+\hat {\mu }+\hat {\nu },\mu }^{(1)}(z_1)
\phi _{m+\hat {\nu },\nu }^{(2)}(z_2)\nonumber \\
&=&\sum _{\mu ',\nu '}W(m|z_1-z_2)^{\mu \nu }_{\mu ',\nu '} 
\phi _{m+\hat {\mu }'+\hat {\nu }',\nu '}^{(2)}(z_2) 
\phi _{m+\hat {\mu }',\mu '}^{(1)}(z_1)
\end{eqnarray}
\noindent where $\phi ^{(i)}$ denote that it act in $i$-th
space $(i=1,2)$.

With the help of the properties of $\phi,\tilde {\phi },\bar {\phi }$
(28,29), we can derive the following relations
from the above face-vertex correspondence relation.

\begin{eqnarray}
&&\tilde {\phi }_{m+\hat {\mu },\mu }^{(1)}(z_1)R_{12}(z_1-z_2)
\phi _{m+\hat {\nu },\nu }^{(2)}(z_2) \nonumber \\
&=&\sum _{\mu '\nu '}W(m|z_1-z_2)^{\mu '\nu }_{\mu \nu '}
\tilde {\phi }_{m+\hat {\mu }'+\hat {\nu },\mu '}^{(1)}(z_1)
\phi _{m+\hat {\mu }+\hat {\nu }',\nu '}^{(2)}(z_2),\\
&&\tilde {\phi }_{m+\hat {\mu }+\hat {\nu },\nu }^{(2)}(z_2)
\tilde {\phi }_{m+\hat {\mu },\mu }^{(1)}(z_1)R_{12}(z_1-z_2)\nonumber \\
&=&\sum _{\mu '\nu '}W(m|z_1-z_2)^{\mu '\nu '}_{\mu \nu }
\tilde {\phi }_{m+\hat {\mu }'+\hat {\nu }',\mu '}^{(1)}
(z_1)\tilde {\phi }_{m+\hat {\nu }',\nu '}^{(2)}(z_2), \\
&&\bar {\phi }_{m,\nu }^{(2)}(z_2)R_{12}(z_1-z_2)\phi _{m,\mu }^{(1)}
(z_1)\nonumber \\
&=&\sum _{\nu '\mu '}W(m-\hat {\mu }-\hat {\nu }'|
z_1-z_2)^{\mu \nu '}_{\mu '\nu }\phi _{m-\hat {\nu },\mu '}^{(1)}(z_1)
\bar {\phi }_{m-\hat {\mu },\nu '}^{(2)}(z_2), \\
&&\bar {\phi }_{m+\hat {\mu }+\hat {\nu },\nu }^{(2)}(z_2)
\bar {\phi }_{m+\hat {\mu },\mu }^{(1)}(z_1)R_{12}(z_1-z_2)\nonumber \\
&=&\sum _{\mu '\nu '}W(m|z_1-z_2)^{\mu '\nu '}_{\mu \nu }
\bar {\phi }_{m+\hat {\mu }'+\hat {\nu }',\mu '}^{(1)}(z_1)
\bar {\phi }_{m+\hat {\nu }',\nu '}^{(2)}(z_2).
\end{eqnarray}

\noindent All of these relations obtained above have described the 
correspondence between face and vertex models. Usually, we call $\phi $
the intertwiner of face-vertex correspondence (Fig.4).

\subsection{Commutation relations for elements of the face boundary
reflecting $k$ matrix}
As mentioned above, in order to obtain a comparatively simple commutation
relations which can be dealed with by algebraic Bethe ansatz method, 
we should change the "vertex"
reflection equations to "face" reflection equations, since where the
new "R-matrix" - face Boltzmann weights have only six non-zero elements
instead of eight non-zero elements in "vertex" case.
For this purpose, by using the three-spin operator $\phi $, 
we change the
vertex boundary reflecting matrix to face boundary reflecting matrix,
and find the commutation relations between the elements of the 
"face" type monodromy matrix, which are useful for
the quantum inverse scattering method.

We first change the matrix $\tilde {K}(z)$ defined in eqs.(16,17) to 
face boundary reflecting matrix $\tilde {K}(m|z)_{\mu }^{\nu }$, using
the unitarity properties of the intertwiner (29), we find
\begin{eqnarray}
\tilde {K}(z)&=&\sum _{\mu }\left\{ \phi _{m,\mu }(-z)
\bar {\phi }_{m,\mu }(-z)\tilde {K}(z)\right. \nonumber \\
&&\times \sum _{\nu }\left. [\phi _{m-\hat {\mu }+\hat {\nu },\nu }(z) 
\tilde {\phi }_{m-\hat {\mu }+\hat {\nu },\nu }(z)]\right\}  \nonumber \\
&=&\sum _{\mu \nu }\phi _{m,\mu }(-z)
\tilde {\phi }_{m-\hat {\mu }+\hat {\nu },\nu }(z)\tilde {K}
(m|z)_{\mu }^{\nu },
\nonumber 
\end{eqnarray}
\noindent where
\begin{eqnarray}
\tilde {K}(m|z)_{\mu }^{\nu }\equiv \bar {\phi }_{m,\mu }(-z)
\tilde {K}(z)\phi _{m-\hat {\mu }+\hat {\nu },\nu }(z)
\end{eqnarray}

Thus, the transfer matrix of eight vertex model with open-boundary 
conditions can be rewritten as

\begin{eqnarray}
t(z)&=&Tr\left( \tilde {K}(z)k(z)\right)\nonumber \\
&=&Tr \left\{ \sum _{\mu \nu }\phi _{m,\mu }(-z)
\tilde {\phi }_{m-\hat {\mu }+\hat {\nu },\nu }(z)
\tilde {K}(m|z)_{\mu }^{\nu }k(z)\right\} \nonumber \\
&=&\sum _{\mu \nu }[\tilde {\phi }_{m-\hat {\mu }+\hat {\nu },\nu }
(z)k(z)\phi _{m,\mu }(-z)]\tilde {K}(m|z)_{\mu }^{\nu }\nonumber \\
&\equiv &\sum _{\mu \nu }k(m|z)^{\mu }_{\nu }\tilde {K}(m|z)_{\mu }^{\nu },
\end{eqnarray}
\noindent which is true for arbitrary $m$.
Here we also introduce the definition of face 
boundary reflecting $k$ matrix as: 
\begin{eqnarray}
k(m|z)_{\nu }^{\mu }
=\tilde {\phi }_{m-\hat {\mu }+\hat {\nu },\nu }(z)k(z)
\phi _{m,\mu }(-z).
\end{eqnarray} 
\noindent We call $m$ and $m-\hat {\mu }+\hat {\nu }$ 
the initial and final weight of 
$k(m|z)_{\nu }^{\mu }$, respectively. Thus, 
we have written out the transfer matrix by using the face form of the model.

Next, we will derive the face reflection equation directly from the 
vertex reflection equation by using the intertwiner. Multiply both
sides of eq.(14) from left by $\tilde {\phi }_{m+\hat {\mu }_0,\mu _0}^{(1)}
(z_1)\tilde {\phi }_{m+\hat {\mu }_0+\hat {\nu }_0,\nu _0}^{(2)}(z_2)$, from
right by $\phi _{m+\hat {\mu }_3,\mu _3}^{(1)}(-z_1)$ 
$\phi _{m+\hat {\mu }_3+\hat {\nu }_3,\nu _3}^{(2)}(-z_2)$, notice the
properties such as $\tilde {\phi }^{(2)}$ commutes with 
$\tilde {\phi }^{(1)}$ and $k_1(z_1)$, use the 
face-vertex correpondence relations.
We can get the face reflection equation (see Appendix A and Fig.5,6) 
[44,46,33],
\begin{eqnarray}
&&k(m+\hat {\mu }_2+\hat {\nu }_1|z_1)_{\mu _1}^{\mu _2}
k(m+\hat {\mu }_3+\hat {\nu }_3|z_2)_{\nu _2}^{\nu _3}\nonumber \\
&&W(m|z_1+z_2)_{\nu _1\mu _2}^{\nu _2\mu _3}
W(m|z_1-z_2)_{\mu _0\nu _0}^{\mu _1\nu _1}\nonumber \\
&=&k(m+\hat {\mu }_1+\hat {\nu }_2|z_2)_{\nu _0}^{\nu _1}
k(m+\hat {\mu }_3+\hat {\nu }_3|z_1)_{\mu _1}^{\mu _2}\nonumber \\
&&W(m|z_1+z_2)_{\mu _0\nu _1}^{\mu _1\nu _2}
W(m|z_1-z_2)_{\nu _2\mu _2}^{\nu _3\mu _3}.
\end{eqnarray}
\noindent Here and below summation 
over repeated indices are assumed. One can find
that this equation is true for arbitrary $\mu _0,\nu _0,\mu _3,\nu _3$.
We should point out here that this face reflection equation is different
from the one proposed by Behrend et al in ref.[44], if the cross-inversion 
relation of face Boltzmann 
weights [41] is applied, the two equations are equivalent. 

Now, we will let the indices in the above relation take special values
so that we can find the necessary commutation relations. When
$\mu _0=\nu _0=0,\mu _3=\nu _3=1$, we get
\begin{eqnarray}
&&k(m|z_1)_0^1k(m-2|z_2)_0^1W(m|z_1+z_2)^{01}_{01}
W(m|z_1-z_2)^{00}_{00}\nonumber \\
&=&k(m|z_2)_0^1k(m-2|z_1)^1_0W(m|z_1+z_2)_{01}^{01}
W(m|z_1-z_2)^{11}_{11}.
\end{eqnarray}
\noindent From the definition of face Boltznamm weights, we know that 
$W(m|z)_{00}^{00}=W(m|z)_{11}^{11}=\frac {h(z+w)}{h(w)}$, so we find 
\begin{eqnarray}
k(m|z_1)_0^1k(m-2|z_2)_0^1=k(m|z_2)_0^1k(m-2|z_1)_0^1.
\end{eqnarray}
\noindent This means that the position of $z_1$ and $z_2$ can be 
exchanged in this form.

Let $\mu _0=\nu _0=\nu _3=0,\mu _3=1$, we obtain
\begin{eqnarray}
&&k(m+2|z_2)_0^0k(m|z_1)_0^1W(m|z_1+z_2)^{00}_{00}
W(m|z_1-z_2)^{01}_{01}\nonumber \\
&=&k(m|z_1)_0^1k(m|z_2)_0^0W(m|z_1+z_2)^{01}_{01}
W(m|z_1-z_2)_{00}^{00}\nonumber \\
&&-k(m|z_2)_0^1k(m|z_1)_0^0W(m|z_1+z_2)_{01}^{01}
W(m|z_1-z_2)^{01}_{10}\nonumber \\
&&-k(m|z_2)_0^1k(m|z_1)_1^1W(m|z_1+z_2)_{01}^{10}W(m|z_1-z_2)_{01}^{01}.
\end{eqnarray}
\noindent Denote
\begin{eqnarray}
A(m|z)&\equiv &k(m|z)_0^0,\nonumber \\
D(m|z)&\equiv &k(m|z)_1^1,\nonumber \\
B(m|z)&\equiv &k(m|z)_0^1,\nonumber \\
C(m|z)&\equiv &k(m|z)_1^0.
\end{eqnarray}
\noindent So, the matrix $\left( \begin{array}{cc} A&B\\ 
C&D\end{array} \right) $ is the
boundary $k$ matrix in the "face" form. Relations (42,43) give the 
commutation relations of $BB$ and $AB$, respectively. In order to use
the algebraic Bethe ansatz method, we must also get the commutation
relation of $DB$.

Let $\mu _0=\mu _3=\nu _3=1,\nu _0=0$, exchange $z_1$ and $z_2$, we get one
relation, exchange $z_1$ and $z_2$ in equation (43), we find another
relation, combine the two relations and with the help of equation (43), we
can find the commutaion relation of $DB$:

\begin{eqnarray}
&&k(m+2|z_2)_1^1k(m|z_1)_0^1W(m+2|z_1+z_2)_{01}^{01}
W(m+2|z_1-z_2)^{10}_{10}\nonumber \\
&&W(m|z_1+z_2)_{00}^{00}W(m|z_1-z_2)_{01}^{01}\nonumber \\
&=&k(m|z_2)_0^1k(m|z_1)_0^0W(m+2|z_1+z_2)_{10}^{01}
W(m|z_1+z_2)_{01}^{01}\nonumber \\
&&\left\{ W(m+2|z_2-z_1)_{11}^{11}W(m|z_2-z_1)_{00}^{00}
-W(m+2|z_2-z_1)_{10}^{01}W(m|z_1-z_2)_{10}^{01}\right\} \nonumber \\
&&+k(m|z_2)_0^1k(m|z_1)_1^1W(m+2|z_2-z_1)_{10}^{01}
W(m|z_2-z_1)_{01}^{01}\nonumber \\
&&\left\{ -W(m+2|z_1+z_2)_{11}^{11}W(m|z_1+z_2)_{00}^{00}
+W(m+2|z_1+z_2)_{10}^{01}W(m|z_1+z_2)_{01}^{10}\right\}\nonumber \\
&&+k(m|z_1)_0^1k(m|z_2)_0^0W(m+2|z_1+z_2)^{01}_{10}
W(m|z_1+z_2)_{01}^{01}\nonumber \\
&&\left\{ -W(m+2|z_2-z_1)_{11}^{11}W(m|z_2-z_1)_{10}^{01}
+W(m+2|z_2-z_1)_{10}^{01}W(m|z_1-z_2)_{00}^{00}\right\}\nonumber \\
&&+k(m|z_1)_0^1k(m|z_2)_1^1W(m+2|z_2-z_1)_{11}^{11}
W(m|z_2-z_1)_{01}^{01}\nonumber \\
&&\left\{ W(m+2|z_1+z_2)_{11}^{11}W(m|z_1+z_2)_{00}^{00}
-W(m+2|z_1+z_2)_{10}^{01}W(m|z_1+z_2)_{01}^{10}\right\} \nonumber \\
\end{eqnarray}
\noindent In the right hand side of this equation, behind $k(m|z_2)_0^1$,
we find not only term $k(m|z_1)_1^1$ but also term $k(m|z_1)_0^0$, this
will cause trouble in the proceeding of the algebraic Bethe ansatz method,
especially in the case where the thermodynamic limit of this system
is taken. In order to solve this problem, we should reformulate 
$A$ and $D$ as $A$ and $\tilde {D}$ so that when we commute $\tilde {D}$
with $B$, only term $\tilde {D}$ exists behind $B(z_2)$ which is
the notation of $k(m|z_2)_0^1$.

It is not easy to find $\tilde {D}$ by direct calculation, but the work
of Sklyanin[16] gives us a hint to formulate the term $\tilde {D}$.
Sklyanin pointed out in his paper that the term $\tilde {D}$ in
six vertex model case is one of the elements of the inverse matrix
of the monodromy matrix. Now we will study the inverse of face form 
monodromy matrix for
the eight vertex model case.

For simplicity, abusing the face weights $m$ etc, 
we can write the face reflection equation as
\begin{eqnarray}
&&W_{12}(z_1-z_2)k_1(z_1)W_{21}(z_1+z_2)k_2(z_2)\nonumber \\
&=&k_2(z_2)W_{12}(z_1+z_2)k_1(z_1)W_{21}(z_1-z_2).
\end{eqnarray}
\noindent It is just the same as the vertex reflection equation. If the 
inverse of $k(z)$ exists, we have

\begin{eqnarray}
&&k_2^{-1}(z_2)W_{12}(z_1-z_2)k_1(z_1)W_{21}(z_1+z_2)\nonumber \\
&=&W_{12}(z_1+z_2)k_1(z_1)W_{21}(z_1-z_2)k_2^{-1}(z_2).
\end{eqnarray}
\noindent That means that $k_2^{-1}(-z_2)$ and $k_2(z_2)$
has similar exchange relation with 
$k_1(z_1)$ in formalism. We have known that the commutation
relation of $k(m+2|z_2)_0^0$ with $k(m|z_1)_0^1$ is comparatively
simple which is also useful for algebraic Bethe ansatz. From the above
equation, we assume that the commutation relation of 
$k^{-1}(m+2|z_2)_0^0$ with $k(m|z_1)_0^1$ should have the similar properties 
in formalism. We know that $k^{-1}_2(z_2)_0^0$ is a linear combination of
$A$ and $D$ in six vertex model. We hope that this is also true for 
the case of eight vertex model. Fortunately, we have obtained the 
expected result.

\noindent Define 
\begin{eqnarray}
Q(m|z-w)_{\mu _3}^{\nu _2}&=&k(m+\hat {\bar {\mu }}_3+
\hat {\nu }_2|z-w)_{\mu _1}^{\mu _2}
W(m|2z-w)_{\nu _1\mu _2}^{\nu _2\bar {\mu }_3}
W(m|-w)_{01}^{\mu _1\nu _1},\nonumber \\
Q'(m|z-w)_{\nu _1}^{\mu _0}&=&
k(m|z-w)_{\mu _1}^{\mu _2}W(m|2z-w)_{\bar {\mu }_0\nu _1}^{\mu _1\nu _2}
W(m|-w)_{\nu _2\mu _2}^{01},
\end{eqnarray}
\noindent where, as usual, summation over repeated indices is assumed and 
$\bar {0}=1,\bar {1}=0$. We can show (Appendix B)
\begin{eqnarray}
Q(m|z-w)_{\nu }^{\nu '}
k(m+\hat {\bar {\nu }}+\hat {\nu }''|z)_{\nu '}^{\nu ''}&=&
\rho (m,\nu |z)\delta _{\nu \nu ''},\nonumber \\
k(m+\hat {\bar {\mu }}''+\hat {\mu }'|z)_{\mu }^{\mu '}
Q'(m|z-w)_{\mu '}^{\mu ''}&=&\rho '(m,\mu |z)\delta _{\mu \mu ''},
\end{eqnarray}
\noindent where $\rho (m,\nu |z)$, and 
$\rho '(m,\mu |z)$ are scalars of 
the "quantum space". We can rescale $Q$ and $Q'$  such that 
$\rho ,\rho '\rightarrow 1$. But this is not necessary in the following 
derivation. Multiply (40) by $Q(m+\hat {\nu }+
\hat {\mu }_0|z_2-w)_{\nu }^{\nu _0}$
from left and by $Q'(m+\hat {\mu }_3+\hat {\nu }'
|z_2-w)_{\nu _3}^{\nu '}$ from right.
Summation over $\nu _0,\nu _3$ gives
\begin{eqnarray}
&&Q(m+\hat {\nu }+\hat {\mu }_0|z_2-w)_{\nu }^{\nu _0}
k(m+\hat {\mu }_3+\hat {\nu }_2|z_1)_{\mu _1}^{\mu _2}
W(m|z_1+z_2)_{\nu _1\mu _2}^{\nu _2\mu _3}\nonumber \\
&&\times W(m|z_1-z_2)_{\mu _0\nu _0}^{\mu _1\nu _1}
\rho '(m+\hat {\mu }_3+\hat {\nu }',\nu '|z_2)\delta _{\nu _2\nu '}
\nonumber \\
&=&\rho (m+\hat {\nu }+\hat {\mu }_0,\nu |z_2)
\delta _{\nu \nu _1}
k(m+\hat{\mu }_3+\hat {\nu }_3|z_1)_{\mu _1}^{\mu _2}
Q'(m+\hat {\mu }_3+\hat {\nu }'|z_2-w)_{\nu _3}^{\nu '}\nonumber \\
&&W(m|z_1+z_2)_{\mu _0\nu _1}^{\mu _1\nu _2}
W(m|z_1-z_2)_{\nu _2\mu _2}^{\nu _3\mu _3}.
\end{eqnarray}
\noindent In the derivation of LHS, we use the fact that 
$W(m|z_1+z_2)_{\nu _1\mu _2}^{\nu _2\mu _3}\not= 0$ only if 
$\hat {\nu _1}+\hat {\mu _2}=\hat {\nu }_2+\hat {\mu _3}$. 
The equation becomes 

\begin{eqnarray}
&&Q(m+\hat {\nu }+\hat {\mu }_0|z_2-w)_{\nu }^{\nu _0}
k(m+\hat {\mu }_3+\hat {\nu }'|z_1)_{\mu _1}^{\mu _2}\nonumber \\
&&W(m|z_1+z_2)_{\nu _1\mu _2}^{\nu '\mu _3}
W(m|z_1-z_2)_{\mu _0\nu _0}^{\mu _1\nu _1}\nonumber \\
&=&k(m+\hat {\mu }_3+\hat {\nu }_3|z_1)_{\mu _1}^{\mu _2}
Q'(m+\hat {\mu }_3+\hat {\nu }'|z_2-w)_{\nu _3}^{\nu '}\nonumber \\ 
&&W(m|z_1+z_2)_{\mu _0\nu }^{\mu _1\nu _2}
W(m|z_1-z_2)_{\nu _2\mu _2}^{\nu _3\mu _3}\nonumber \\
&&\times \frac {\rho (m+\hat {\nu }+\hat {\mu }_0,\nu |z_2)}
{\rho '(m+\hat {\mu }_3+\hat {\nu }',\nu '|z_2)},
\end{eqnarray}
\noindent which is similar to (40). Put $\mu _0=\nu =\nu '=0,
\mu _3=1$, and notice $m+\hat {0}+\hat {0}=m+2,m+\hat {0}+\hat {1}=m$.
We then have 
\begin{eqnarray}
&&Q(m+2|z_2-w)_0^1k(m|z_1)_0^0W(m|z_1+z_2)_{10}^{01}W(m|z_1-z_2)_{01}^{01}
\nonumber \\
&&+Q(m+2|z_2-w)_0^0k(m|z_1)_0^1W(m|z_1+z_2)_{01}^{01}W(m|z_1-z_2)_{00}^{00}
\nonumber \\
&&+Q(m+2|z_2-w)_0^1k(m|z_1)_1^1W(m|z_1+z_2)_{01}^{01}W(m|z_1-z_2)_{01}^{10}
\nonumber \\
&=&k(m|z_1)_0^1Q'(m|z_2-w)_0^0W(m|z_1+z_2)_{00}^{00}W(m|z_1-z_2)_{01}^{01}
\frac {\rho (m+2,0|z_2)}{\rho '(m,0|z_2)}.
\nonumber \\
\end{eqnarray}
\noindent Using the same derivation as that in Appendix B, we can calculate
\begin{eqnarray}
\frac {\rho (m+2,0|z_2)}{\rho '(m,0|z_2)}=
\frac {(-1)h(wa)h(w(a+1))}{h(w(a-1))h(w(a+2))},
\end{eqnarray}
\noindent where $a\equiv m+\gamma $ and $\gamma $ is a parameter in 
defining $\phi $. From (48) and considering the explicit form of $W$, 
we can write $Q,Q'$ by components of $k(z)$,
\begin{eqnarray}
Q(m|z-w)_0^1=k(m-2|z-w)_0^1W(m|2z-w)_{11}^{11}W(m|-w)_{01}^{01},
\end{eqnarray}
\noindent which is proportional to $B(m-2|z-w)$, and
\begin{eqnarray}
Q(m|z-w)_0^0&=&-\left[ k(m|z-w)_1^1W(m|2z-w)_{10}^{10}\right. \nonumber \\
&&\left. -k(m|z-w)_0^0W(m|2z-w)_{10}^{01}
\right] W(m|-w)_{01}^{01}\nonumber \\
&=&-Q'(m|z-w)_0^0\nonumber \\
&\equiv &-\tilde {D}(m|z-w)W(m|-w)_{01}^{01},
\end{eqnarray}
\noindent where 
\begin{eqnarray}
\tilde {D}(m|z)&\equiv &-k(m|z)_0^0W(m|2z+w)_{10}^{01}
+k(m|z)_1^1W(m|2z+w)_{10}^{10}\nonumber \\
&=&\frac {-Q(m|z)}{W(m|-w)_{01}^{01}}
\end{eqnarray}
\noindent is indeed a linear combination of $A(m|z)\equiv k(m|z)_0^0$
and $D(m|z)\equiv k(m|z)_1^1$. Subsititute (53-55) into (52)
and write $k(m|z_2-w)_1^1$ in the equation in terms of 
$\tilde {D}(m|z_2-w)$ and $A(m|z_2-w)$. We then change the  
parameter $z_2-w$ to $z_2$ giving the expected equation
\begin{eqnarray}
&&\tilde {D}(m+2|z_2)B(m|z_1)\nonumber \\
&=&B(m|z_1)\tilde {D}(m|z_2)\frac {h(z_1-z_2-w)h(z_1+z_2+2w)}
{h(z_1+z_2+w)h(z_1-z_2)}\nonumber \\
&&+B(m|z_2)A(m|z_1)\frac {h(2z_1)h(z_1+z_2+w(a+2))h(2z_2+2w)}
{h(2z_1+w)h(z_1+z_2+w)h(w(a+1))}\nonumber \\
&&+B(m|z_2)\tilde {D}(m|z_1)
\frac {h(z_2-z_1+w(a+1))h(2z_2+2w)h(w)}
{h(2z_1+w)h(z_1-z_2)h(w(a+1))}.
\end{eqnarray}
\noindent where $h(z)\equiv \sigma _0(z)$. In the derivation we have
used a formular of $\theta $ function [7]
\begin{eqnarray}
&&h(u+x)h(u-x)h(v+y)h(v-y)-h(u+y)h(u-y)h(v+x)h(v-x)\nonumber \\
&=&h(u+v)h(u-v)h(x+y)h(x-y).
\end{eqnarray}

We need also to change $D$ in the exchange relation of $A$ and $B$ by
linear combination of $\tilde {D}$ and $A$. Thus (43) is rewritten
as
\begin{eqnarray}
&&A(m+2|z_2)B(m|z_1)\nonumber \\
&=&B(m|z_1)A(m|z_2)\frac {h(z_1+z_2)h(z_1-z_2+w)}
{h(z_1+z_2+w)h(z_1-z_2)}\nonumber \\
&&-B(m|z_2)A(m|z_1)\frac {h(2z_1)h(z_1-z_2+w(a+1))h(w)}
{h(2z_1+w)h(z_1-z_2)h(w(a+1))}\nonumber \\
&&-B(m|z_2)\tilde {D}(m|z_1)\frac {h(-z_1-z_2+wa)[h(w)]^2}
{h(z_1+z_2+w)h(2z_1+w)h(w(a+1))}.
\end{eqnarray}
\noindent We have also used (58) in the derivation. With 
the permutation relations of $AB$, $\tilde {D}B$ and $BB$ (57,59,42), 
we can obtain the algebraic
Bethe ansatz equations provided

\noindent 1.The transfer matrix $t(z)$ is a linear combination of $A$ and
$\tilde {D}$,

\noindent 2.There is a "vaccum state" of the quantum space which is an
eigenstate of $A$ and $\tilde {D}$ but not an eigenstate of $B$.

\noindent We will study these problems in section 4.

\section{Vaccum state and boundary conditions}
\subsection{Vaccum state}
The algebraic Bethe ansatz requires to construct a state of the "quantum"
space , which is an eigenstate of operators $A$ and $D$ with all 
spectrum parameter $z$. This state is called a vaccum state, which is
also an eigenstate of $\tilde {D}$. Before introducing the vaccum state,
let us make some preperations. We first expresss $S$ 
and $T$ (see Eq.(18)) by the "face laguage", change their auxiliary 
indices to face indices, and express $k(m|z)_{\mu }^{\nu }$ by such
expressions of $S$ and $T$. From (23,39), 
the operator $k(z)$ with "face" indices can 
be written as
\begin{eqnarray}
k(m|z)_{\mu '}^{\nu '}=\tilde {\phi }_{m+\hat {\mu }'-\hat {\nu }',\mu '}
(z)T(z)K(z)S(z)\phi _{m,\nu '}(-z).
\end{eqnarray}
\noindent From (28,29) we have 
\begin{eqnarray}
K(z)=\sum _{\mu \nu }\left\{ 
\phi _{m_0+\hat {\mu }-\hat {\nu },\mu }(z)
\tilde {\phi }_{m_0+\hat {\mu }-\hat {\nu },\mu }(z)K(z)
\phi _{m_0,\nu }(-z)\bar {\phi }_{m_0,\nu }(-z)\right\} . 
\end{eqnarray}
\noindent Combining these two equations gives
\begin{eqnarray}
k(m|z)_{\mu '}^{\nu '}=\sum _{\mu \nu }K(m_0|z)^{\nu }_{\mu }
T(m-\hat {\nu }',m_0-\hat {\nu }|z)_{\mu '\mu }
S(m,m_0|z)_{\nu '\nu },
\end{eqnarray}
\noindent where we define
\begin{eqnarray}
K(m_0|z)_{\mu }^{\nu }&\equiv &\tilde {\phi }_{m_0+
\hat {\mu }-\hat {\nu },\mu }(z)K(z)\phi _{m_0 ,\nu }(-z),\\
T(m-\hat {\nu }',m_0-\hat {\nu }|z)_{\mu '\mu }&\equiv &
\tilde {\phi }_{m+\hat {\mu }'-\hat {\nu }',\mu '}(z)T(z)
\phi _{m_0+\hat {\mu }-\hat {\nu },\mu }(z),\\
S(m,m_0|z)_{\nu '\nu }&\equiv &\bar {\phi }_{m_0,\nu }(-z)S(z)
\phi _{m,\nu '}(-z).
\end{eqnarray}
\noindent Equation (62) is true for all $m$ and $m_0$.

Next, assume that we can properly choose the parameters of the right
reflecting matrix such that for given $m_0$ and all $z$, 
$K(m_0|z)_1^0=0$. This is possible. We will study this problem in the
second part of this section. Actually, this requirement constitute the
only restriction on the boundary matrices in our approach.

Then, multiply $\tilde {\phi }_{m,1}^{(1)}(z)
\bar {\phi }_{m_0+1,0}^{(2)}(-z)$ from left and multiply 
$\phi _{m_0+1,0}^{(1)}(z)\phi _{m,1}^{(2)}(-z)$ from right to 
equation (21). Using face-vertex correspondence relations
{32-36) we can prove the following exchange relation
of $S$ and $T$ with face indices,
\begin{eqnarray}
&&W(m+1|2z)_{11}^{11}S(m-1,m_0+1|z)_{10}T(m,m_0|z)_{10}\nonumber \\
&&+W(m+1|2z)_{10}^{01}S(m+1,m_0+1|z)_{00}T(m,m_0|z)_{00}\nonumber \\
&=&W(m_0+1|2z)_{10}^{01}T(m+1,m_0+1|z)_{11}S(m,m_0|z)_{11}\nonumber \\
&&+W(m_0-1|2z)_{00}^{00}T(m+1,m_0-1|z)_{10}S(mm_0|z)_{10},
\end{eqnarray}
\noindent which will be useful in our derivation. For later convenience we 
rewrite $S,T$ as
\begin{eqnarray}
S(z)=R_{l0}(u_l+z)R_{l-1,0}(u_{l-1}+z)\cdots R_{10}(u_1+z),\nonumber \\
T(z)=R_{01}(z-u_1)R_{02}(z-u_2)\cdots R_{0l}(z-u_l).
\end{eqnarray}
\noindent They are acting on the space $V_0\otimes V_1\otimes \cdots
\otimes V_l$, where $V_0$ is the auxiliary space. 
The final preparation is the following observation. 
In Eqs.(32-36), the summation over face indices has at most two terms in eight
vertex model. In the following cases, due to the non zero condition of
$W(m|z)^{\mu '\nu '}_{\mu \nu }$, there is actually only one term in RHS of
the equations,

\begin{eqnarray}
\begin{array}{ll}
1.~~Eq.(32) &when~~\mu =\nu \\
2.~~Eq.(33) &when~~\mu \not =\nu \\
3.~~Eq.(35) &when~~\mu \not =\nu .
\end{array}
\end{eqnarray}

\noindent Thus we have (Fig.7)
\begin{eqnarray}
&&R_{12}(z_1-z_2)\phi _{m+2,0}^{(1)}(z_1)\phi _{m+1,0}^{(2)}(z_2)\nonumber \\
&=&W(~~|z_1-z_2)_{00}^{00}\phi _{m+2,0}^{(2)}(z_2)
\phi _{m+1,0}^{(1)}(z_1),
\end{eqnarray}

\begin{eqnarray}
&&\tilde {\phi }_{m-1,1}^{(1)}(z_1)R_{12}(z_1-z_2)
\phi _{m+1,0}^{(2)}(z_2)\nonumber \\
&=&W(m|z_1-z_2)_{10}^{10}\tilde {\phi }_{m,1}^{(1)}(z_1)
\phi _{m,0}^{(2)}(z_2),
\end{eqnarray}

\begin{eqnarray}
&&\bar {\phi }_{m,1}^{(2)}(z_2)R_{12}(z_1-z_2)\phi _{m,0}^{(1)}(z_1)\nonumber \\
&=&W(m|z_1-z_2)_{01}^{01}\phi _{m+1,0}^{(1)}(z_1)
\bar {\phi }_{m-1,1}^{(2)}(z_2),
\end{eqnarray}

\noindent which will be repeatedly used in the proof of vaccum state. 

The above are preparations for introducing the vaccum state. Now, define 
the vaccum state as
\begin{eqnarray}
|0>^m_{m_0}\equiv \phi _{m_0,0}^{(l)}(u_l)
\phi _{m_0-1,0}^{(l-1)}(u_{l-1})\cdots \phi _{m_0-(l-2),0}^{(2)}(u_2)
\phi _{m_0-(l-1),0}^{(1)}(u_1)
\end{eqnarray}
\noindent where $m=m_0-l$. This is precisely the same vaccum state 
introduced by Baxter in the original work of Bethe ansatz for eight
vertex model with periodic boundary conditions [3]. For the
vaccum state defined in (72) we can show that $S(m,m_0|z)_{00}$ and
$T(m,m_0|z)_{11}$ change $|0>^m_{m_0}$ to $|0>^{m-1}_{m_0-1}$,
while $S(m,m_0|z)_{11}$ and $T(m,m_0|z)_{00}$ change $|0>^m_{m_0}$
to $|0>_{m_0+1}^{m+1}$ (with some coefficients). The operators
$S(m,m_0|z)_{01}$ and $T(m,m_0|z)_{10}$ change $|0>_{m_0}^m$ to zero. 
Following is the proof.

We have
\begin{eqnarray}
&&S(m,m_0|z)_{00}\nonumber \\
&=&\bar {\phi }_{m_0,0}^{(0)}(-z)S(z)
\phi _{m,0}^{(0)}(-z)\nonumber \\
&=&\bar {\phi }_{m_0,0}^{(0)}(-z)R_{l0}(u_l+z)R_{l-1,0}(u_{l-1}+z)
\cdots R_{1,0}(u_1+z)\phi _{m,0}^{(0)}(-z_1). 
\end{eqnarray}
\noindent Since vectors and operators belonging to different spaces may
change their positions in an equation, we obtain
\begin{eqnarray}
&&S(m,m_0|z)_{00}|0>_{m_0}^m\nonumber \\
&=&\bar {\phi }_{m_0,0}^{(0)}(-z)
R_{l0}(u_l+z)\phi _{m_0,0}^{(l)}(u_l)R_{l-1,0}(u_{l-1}+z)
\phi _{m_0-1,0}^{(l-1)}(u_{l-1})\nonumber \\
&&\cdots R_{20}(u_2+z)\phi _{m_0-(l-2),0}^{(2)}(u_2)
R_{10}(u_1+z)\phi _{m+1,0}^{(1)}(u_1)\phi _{m,0}^{(0)}(-z).
\end{eqnarray}
\noindent By using (69), we move $\phi _{~~,0}^{(0)}(-z)$ towards left
step by step. At each step we eliminate an R matrix getting a $W$
factor, and change the face weight of $\phi _{~~,0}^{(0)}(-z)$ and 
that of $\phi _{~~,0}^{(i)}$ encountered. Thus we have
\begin{eqnarray}
S(m,m_0|z)_{00}|0>_{m_0}^m&=&\cdots R_{20}(u_2+z)
\phi _{m+2,0}^{(2)}(u_2)
\phi _{m+1,0}^{(0)}(-z)\phi _{m,0}^{(1)}(u_1) \nonumber \\
&&\times W(~~|u_1+z)_{00}^{00} \nonumber \\
&=&\cdots \nonumber \\
&=&\bar {\phi }_{m_0,0}^{(0)}(-z)\phi _{m_0,0}^{(0)}(-z)
\phi _{m_0-1,0}^{(l)}(u_l)\cdots \phi_{m,0}^{(1)}(u_1) \nonumber \\
&&\times \prod _{i=1}^lW(~~|u_i+z)_{00}^{00}.
\end{eqnarray}

\noindent Due to the orthogonal relation (28),this becomes
\begin{eqnarray}
S(m,m_0|z)_{00}|0>_{m_0}^m&=&
\prod _{i=1}^l W(~~|u_i+z)_{00}^{00}|0>_{m_0-1}^{m-1}\nonumber \\
&\equiv &s_{00}(z)|0>_{m_0-1}^{m-1}.
\end{eqnarray}
\noindent Similarly, one can show
\begin{eqnarray}
S(m,m_0|z)_{01}|0>_{m_0}^m&=&\bar {\phi }_{m_0,1}^{(0)}(-z)
\phi _{m_0,0}^{(0)}(-z)\cdots \nonumber \\ 
&=&0.
\end{eqnarray}
\noindent For the action of $S_{11}$, from (65,67) and (72), we write
\begin{eqnarray}
&&S(m,m_0|z)_{11}|0>_{m_0}^m\nonumber \\
&=&\bar {\phi }_{m_0,1}^{(0)}(-z)
R_{l0}(u_l+z)\phi _{m_0,0}^{(l)}(u_l)R_{l-1,0}(u_{l-1}+z)
\phi _{m_0-1,0}^{(l-1)}(u_{l-1})\nonumber \\
&&\cdots R_{10}(u_1+z)\phi _{m+1,0}^{(0)}(u_1)\phi _{m,1}^{(0)}(-z).
\end{eqnarray}

\noindent Using (71) we move $\bar {\phi }_{~~,1}^{(0)}(-z)$ towards 
right step by step. At each step we eliminate an R matrix getting a 
$W$ factor, and change the face weight of 
$\bar {\phi }_{~~,1}^{(0)}(-z)$ and $\phi _{~~,0}^{(i)}$ encountered. 
We then have (Fig.8)
\begin{eqnarray}
S(m,m_0|z)_{11}|0>_{m_0}^m&=&\prod _{i=1}^lW(m_0-(i-1)|u_i+z)_{01}^{01}
\bar {\phi }_{m_0-l,1}^{(0)}(-z) \nonumber \\
&&\times \phi _{m,1}^{(0)}(-z) \phi _{m_0+1,0}^{(l)}(u_l)\cdots 
\phi _{m+2,0}^{(1)}(u_1)\nonumber \\
&=&\prod _{i=1}^lW(m_0-(i-1)|u_i+z)_{01}^{01}|0>_{m_0+1}^{m+1}\nonumber \\
&\equiv &s_{11}(m|z)|0>_{m_0+1}^{m+1},
\end{eqnarray}
\noindent here $m_0-l=m$. Similar derivation 
can be performed for $T$, giving
\begin{eqnarray}
T(m,m_0|z)_{00}|0>_{m_0}^m&=&\prod _{i=1}^lW(~~|z-u_i)_{00}^{00}
|0>_{m_0+1}^{m+1}\nonumber \\
&\equiv &t_{00}(z)|0>_{m_0+1}^{m+1},\nonumber \\
T(m,m_0|z)_{10}|0>_{m_0}^m&=&0,\\
T(m,m_0|z)_{11}|0>_{m_0}^m&=&\prod _{i=1}^lW(m+(i-1)|z-u_i)_{10}^{10}
|0>_{m_0-1}^{m-1}\nonumber \\
&\equiv &t_{11}(m|z)|0>_{m_0-1}^{m-1},
\end{eqnarray}
\noindent which completes our proof.
 
We now assume that in (62), $K(m_0|z)_1^0=0$, and consider 
$\hat {\mu }\equiv (-1)^{\mu }$, obtaining
\begin{eqnarray}
k(m|z)_0^0&=&K(m_0|z)_0^0T(m-1,m_0-1|z)_{00}S(m,m_0|z)_{00}\nonumber \\
&&+K(m_0|z)_0^1T(m-1,m_0+1|z)_{00}S(m,m_0|z)_{01}\nonumber \\
&&+K(m_0|z)_1^1T(m-1,m_0+1|z)_{01}S(m,m_0|z)_{01},\\
k(m|z)_1^1&=&K(m_0|z)_0^0T(m+1,m_0-1|z)_{10}S(m,m_0|z)_{10}\nonumber \\
&&+K(m_0|z)_0^1T(m+1,m_0+1|z)_{10}S(m,m_0|z)_{11}\nonumber \\
&&+K(m_0|z)_1^1T(m+1,m_0+1|z)_{11}S(m,m_0|z)_{11}.
\end{eqnarray}
\noindent By (66), $k(m|z)_1^1$ can be written as
\begin{eqnarray}
k(m|z)_1^1&=&\frac {K(m_0|z)_0^0}{W(m_0-1|2z)_{00}^{00}}\left\{
W(m+1|2z)_{11}^{11}S(m-1,m_0+1|z)_{10}T(m,m_0|z)_{10}\right.\nonumber \\
&&\left. +W(m+1|2z)_{10}^{01}S(m+1,m_0+1|z)_{00}
T(m,m_0|z)_{00}\right\}\nonumber \\
&&+K(m_0|z)_0^1T(m+1,m_0+1|z)_{10}S(m,m_0|z)_{11}\nonumber \\
&&+\left\{ K(m_0|z)_1^1-\frac {K(m_0|z)^0_0
W(m_0+1|2z)_{10}^{01}}{W(m_0-1|2z)_{00}^{00}}\right\} \nonumber \\
&&\times T(m+1,m_0+1|z)_{11}S(m,m_0|z)_{11}.
\end{eqnarray}
\noindent Acting on the vaccum state $|0>_{m_0}^m$ and by equations 
(76,79-81), 
these two operators yield
\begin{eqnarray}
k(m|z)_0^0|0>_{m_0}^m&=&K(m_0|z)_0^0T(m-1,m_0-1|z)_{00}
s_{00}(z)|0>_{m_0-1}^{m-1}+0+0\nonumber \\
&=&K(m_0|z)_0^0t_{00}(z)s_{00}(z)|0>_{m_0}^m\nonumber \\
&\equiv &\tau (m|z)_0^0|0>_{m_0}^m,
\end{eqnarray}
\noindent and
\begin{eqnarray}
k(m|z)_1^1|0>_{m_0}^m&=&\frac {K(m_0|z)_0^0W(m+1|2z)_{10}^{01}}
{W(m_0-1|2z)_{00}^{00}}s_{00}(z)t_{00}(z)|0>_{m_0}^m+0 \nonumber \\
&&+\left\{ K(m_0|z)_1^1
-\frac {K(m_0|z)_0^0W(m_0+1|2z)_{10}^{01}}{W(m_0-1|2z)_{00}^{00}}\right\}
\nonumber \\
&&\times t_{11}(m+1|z)s_{11}(m|z)|0>_{m_0}^m\nonumber \\
&\equiv &\tau (m|z)_1^1|0>_{m_0}^m.
\end{eqnarray}
\noindent From equations (85,86) we see that when $K(m_0|z)_1^0=0$, the
vaccum state $|0>_{m_0}^m$ is indeed an eigenstate of 
$A=k(m|z)_0^0$ and $D=k(m|z)_1^1$. Thus it is also an eigenstate of 
$\tilde {D}$. The eigenvalues depend on $m, \left\{ u_i \right\} ,l$  
and $z$, the spectrum of $A$ and $D$.

\subsection{Boundary conditions}
The algebraic Bethe ansatz requires a vaccum state, which needs the 
right boundary to satisfy 
\begin{eqnarray}
K(m_0|z)_1^0=0.                                 
\end{eqnarray}
\noindent Also, the transfer matrix $t(z)$ must be a linear combination
of $A$ and $\tilde {D}$, which impose the left boundary to satisfy
\begin{eqnarray}
\tilde {K}(m^0|z)_0^1=0
\end{eqnarray}
\noindent and
\begin{eqnarray}
\tilde {K}(m^0|z)_1^0=0,
\end{eqnarray}
\noindent for $m^0$ which we will specify in the next section. We will
see that $m^0$ is constrained with $m,m_0$ of vaccum state by
\begin{eqnarray}
&&m+l\hat {0}=m+l=m_0,\nonumber \\
&&m+(-\hat {1}+\hat {0})l'=m+2l'=m^0.
\end{eqnarray}
\noindent In equation (90), $l$ is the column number of the lattice and 
$l'$ is a positive integer, which will be the number of $B$ operators in 
constructing the eigenstates of $t(z)$ (see (105)).

From the general solution of RE (15,17),
\begin{eqnarray}
K(z)&=&\sum _{\alpha }C_{\alpha }\frac {I_{\alpha }}
{\sigma _{\alpha }(-z)},\nonumber \\
\tilde {K}(z)&=&\sum _{\alpha }\tilde {C}_{\alpha }
\frac {I_{\alpha }}{\sigma _{\alpha }(z+w)},\nonumber
\end{eqnarray}
\noindent and the definitions
\begin{eqnarray}
K(m_0|z)_{\mu }^{\nu }&\equiv &
\tilde {\phi }_{m_0+\hat {\mu }-\hat {\nu },\mu }(z)K(z)
\phi _{m_0,\nu }(-z),\nonumber \\
\tilde {K}(m|z)_{\mu }^{\nu }&\equiv &\bar {\phi }_{m,\mu }(-z)
\tilde {K}(z)\phi _{m-\hat {\mu }+\hat {\nu },\nu}(z),
\end{eqnarray}
\noindent we can derive the following results (Appendix C),
\begin{eqnarray}
K(m|z)_1^0&=&(-1)\frac {\sum _{\alpha }C_{\alpha }(-1)^{\alpha _2}
\sigma _{\alpha }(wa+w\beta -{1\over 2})}{h(z+w\beta +w-{1\over 2})
h(wa-w)},\\
\tilde {K}(m|z)_1^0&=&\frac {\sum _{\alpha }\tilde {C}_{\alpha }
(-1)^{\alpha _2}\sigma _{\alpha }(-w(a-1)+w\beta -{1\over 2})}
{h(-z+w\beta -{1\over 2})h(wa)},\\
\tilde {K}(m|z)_0^1&=&(-1)\frac {\sum _{\alpha }
\tilde {C}_{\alpha }(-1)^{\alpha _2}\sigma _{\alpha }(w(a+1)+w\beta
-{1\over 2})}{h(-z+w\beta -{1\over 2})h(wa)},
\end{eqnarray}
\noindent where $a\equiv m+\gamma $. We can easily see from (92-94) that 
those conditions (87-89) are actually independent of $z$, and depend 
only on $\left\{ C_{\alpha }\right\},
\left\{ \tilde {C}_{\alpha }\right\} $ and parameters $\beta ,\gamma ,w,
\tau $,etc.. For any given generic $\left\{ \tilde {C}_{\alpha }\right\} $,
we may solve the equation
\begin{eqnarray}
\sum _{\alpha }\tilde {C}_{\alpha }(-1)^{\alpha _2}
\sigma _{\alpha }(2\eta )=0.
\end{eqnarray}
\noindent The LHS of (95) is a doubly quasi-periodic holomorphic 
function of $\eta $. From the quasi-periodicity we see that [7,40,41] it
has four zeros in $\Lambda _{\tau }:\eta \rightarrow \eta +1,\eta 
\rightarrow \eta +\tau $. Assume $\eta _1,\eta _2$ are two different
zeros, we obtain $\alpha $ and $\beta $ by solving
\begin{eqnarray}
-w(a-1)+w\beta -{1\over 2}&=&2\eta _1,\nonumber \\
w(a+1)+w\beta -{1\over 2}&=&2\eta _2.
\end{eqnarray}
\noindent Then from $a=m^0+\gamma $ we obtain $\gamma $ according to a
given $m^0$. Using such $\beta ,\gamma $ from generic 
$\left\{ \tilde {C}_{\alpha }\right\} $, one may construct
$\phi ,\bar {\phi }$ and $\tilde {\phi }$ with which the equations 
(88) and (89) are satisfied. Since $\beta $ and $\gamma $ 
are completely determined, equation (87) is a constraint for 
$\left\{ C_{\alpha } \right\} $ at the right boundary. This is the 
only restriction we must impose for our approach. There are 3 free
parameters at the right boundary for general solution (15),
(an overall scalar of $K$ is not important). With the constraint (87),
there are still 2 free parameters left. If we further require 
\begin{eqnarray}
K(m_0|z)_0^1=0,
\end{eqnarray}
\noindent then only one free parameter at right boundary survives. 
We can show 
\begin{eqnarray}
\frac {K(m_0|z)^1_1}{K(m_0|z)^0_0}
=\frac {h(\xi -z)h(aw+\xi +z)}{h(\xi +z)h(aw+\xi -z)}
\nonumber
\end{eqnarray}
\noindent where $\xi $ is the right boundary free parameter
introduced in [44], and similarly for the left boundary.
In this case the left and right boundary are equivalent to
that of SOS model (except a symmetric factor) introduced 
by Behrend, Pearce and Brien in [44] where they derive the 
solutions directly from the face reflection equation. This implies that
our approach can be used for the Bethe ansatz of SOS model with such
boundary conditions. 

\section{Bethe ansatz}
We see that for a given $m^0$, properly choosing $\beta $ and $\gamma $, we 
can ensure 
\begin{eqnarray}
\tilde {K}(m^0|z)_0^1=\tilde {K}(m^0|z)_1^0=0
\end{eqnarray}
\noindent at the left boundary. We need the parameters 
$\left\{ C_{\alpha } \right\} $ at the right boundary to satisfy
\begin{eqnarray}
\sum _{\alpha }C_{\alpha }(-1)^{\alpha _2}
\sigma _{\alpha }(wa+w\beta -{1\over 2})=0
\end{eqnarray}
\noindent for an integer $m_0$, where $a\equiv m_0+\gamma $, to
ensure $K(m_0|z)_1^0=0$. Assume that $2l'=m^0-(m_0-l)$ is a positive
even integer, where $l$ is the number of column of the lattice. We can
proceed the standard algebraic Bethe ansatz [3,4,7,16] as following.

We have the transfer matrix from Eq.(38),
\begin{eqnarray}
t(z)&=&\tilde {K}(m^0|z)_0^0k(m^0|z)_0^0\nonumber \\
&&+\tilde {K}(m^0|z)_1^1k(m^0|z)_1^1.
\end{eqnarray}
\noindent Due to (44) and (56), $t(z)$ can be rewritten as
\begin{eqnarray}
t(z)=\mu _0(z)A(m^0|z)+\mu _1(z)\tilde {D}(m^0|z)
\end{eqnarray}
\noindent We also have the exchange relations of $A,\tilde {D}$ with
$B$, Eq.(57) and (59). They can be compactly written as
\begin{eqnarray}
A(m|u)B(m-2|v)&=&a_{00}(m,u,v)B(m-2|v)A(m-2|u)\nonumber \\
&&+b_{00}(m,u,v)B(m-2|u)A(m-2|v)\nonumber \\
&&+b_{01}(m,u,v)B(m-2|u)\tilde {D}(m-2|v),\\
\tilde {D}(m|u)B(m-2|v)&=&a_{11}(m,u,v)B(m-2|v)\tilde {D}(m-2|u)\nonumber \\
&&+b_{10}(m,u,v)B(m-2|u)A(m-2|v)\nonumber \\
&&+b_{11}(m,u,v)B(m-2|u)\tilde {D}(m-2|v).
\end{eqnarray}
\noindent Besides, we have (42), the exchange relation of $B$'s,
\begin{eqnarray}
B(m|u)B(m-2|v)=B(m|v)B(m-2|u).
\end{eqnarray}
\noindent Consider a vector of the quantum space
\begin{eqnarray}
\Phi =B(m^0-2|z_1)B(m^0-4|z_2)\cdots 
B(m^0-2l'|z_{l'})|0>_{m_0}^m,
\end{eqnarray}
\noindent where the number of $B$ is $l'$, $m^0-2l'=m$. 
Due to (104), it is symmetric
for $z_i$'s. We will show that for properly chosen 
$z_1,\cdots ,z_{l'}$ which satisfy the so called
Bethe ansatz equations, $\Phi $ is an eigenvector (eigenstate) of $t(z)$.

We have
\begin{eqnarray}
&&t(z)\Phi \nonumber \\
&=&\left\{ \mu _0(z)A(m^0|z)+\mu _1(z)\tilde {D}(m^0|z)
\right\} \Phi \nonumber \\
&=&\left\{ \mu _0(z)A(m^0|z)+\mu _1(z)\tilde {D}(m^0|z)\right\} \nonumber \\
&&\times B(m^0-2|z_1)\cdots B(m|z_{l'})|0>_{m_0}^m \nonumber \\
&=&\left\{ B(m^0-2|z_1)\left[ \mu _0(z)a'_{00}A(m^0-2|z)
+\mu _1(z)a'_{11}\tilde {D}(m^0-2|z)\right] \right. \nonumber \\
&&+B(m^0-2|z)\left[ \mu _0(z)b'_{00}A(m^0-2|z_1)
+\mu _0(z)b'_{01}\tilde {D}(m^0-2|z_1)\right. \nonumber \\
&&\left. \left. +\mu _1(z)b'_{10}A(m^0-2|z_1)+\mu _1(z)b'_{11}
\tilde {D}(m^0-2|z_1)\right] \right\} \nonumber \\
&&\times B(m^0-4|z_2)\cdots |0>_{m_0}^m \nonumber \\
&=&\left\{ B(m^0-2|z_1)\left[ (\mu a')_0A(m^0-2|z)+(\mu a')_1
\tilde {D}(m^0-2|z)\right] \right. \nonumber \\
&&\left. +B(m^0-2|z) \left[ (\mu b')_0A(m^0-2|z_1)+(\mu b')_1
\tilde {D}(m^0-2|z_1)\right] \right\} \nonumber \\
&&\times B(m^0-4|z_2)\cdots |0>_{m_0}^{m} \nonumber \\
&=&\left\{ B(m^0-2|z_1)B(m^0-4|z_2)\left[ (\mu a'a'')_0
A(m^0-4|z)\right. \right. \nonumber \\
&&\left. +(\mu a'a'')_1\tilde {D}(m^0-4|z) \right] \nonumber \\
&&+B(m^0-2|z)B(m^0-4|z_2)\left[ (\mu b'a_1'')_0A(m^0-4|z_1) \right. 
\nonumber \\
&&\left. +(\mu b'a_1'')\tilde {D}(m^0-4|z_1)\right] \nonumber \\
&&+B(m^0-2|z)B(m^0-4|z_1)\left[ (\mu b'b'')_0A(m^0-4|z_2)
\right. \nonumber \\
&&\left. \left. +(\mu b'b'')_1\tilde {D}(m^0-4|z_2)\right] \right\}
\nonumber \\
&&\times B(m^0-6|z_3)\cdots |0>_{m_0}^m ,
\end{eqnarray}
\noindent where $a',a'',a_1'',b',b''$ are $2\times 2$ matrices. The 
matrices $a',a'',a_1''$ are diagonal. The elements of these matrices are
determined by $m^0,z$ and $z_i$ via Eqs.(102,103). The notations of 
these matrices are 
\begin{eqnarray}
a(m^0,z,z_1)&\equiv &a',\nonumber \\
a(m^0-2,z,z_2)&\equiv &a'',\nonumber \\
a(m^0-2,z_1,z_2)&\equiv &a_1'',\nonumber \\
b(m^0,z,z_1)&\equiv &b',\nonumber \\
b(m^0-2,z,z_2)&\equiv &b''
\end{eqnarray}
\noindent for short. The notations $(\mu \cdots )_i$ 
represents the $i$-th component
of the product of the row vector $\mu $ with matrix $"\cdots "$. In the
following derivation, the notations are similar. Repeatedly using
Eqs.(102,103) to move $A$ and $\tilde {D}$ to the right of all $B$'s, we 
obtain
\begin{eqnarray}
t(z)\Phi &=&B(m^0-2|z_1)\cdots B(m^0-2l'|z_{l'})\nonumber \\
&&\times \left[ (\mu a'a''\cdots a^{(l')})_0
A(m^0-2l'|z)+(\mu a'a''\cdots a^{(l')})_1
\tilde {D}(m^0-2l'|z)\right] |0>^m_{m_0}\nonumber \\
&&+B(m^0-2|z)B(m^0-4|z_2)\cdots B(m^0-2l'|z_{l'})\nonumber \\
&&\times \left[ 
(\mu b'a_1''\cdots a_1^{(l')})_0A(m^0-2l'|z_1)\right. \nonumber \\
&&\left. +(\mu b'a_1''\cdots a_1^{(l')})_1\tilde {D}(m^0-2l'|
z_1)\right] |0>_{m_0}^m \nonumber\\
&&+B(m^0-2|z)B(m^0-4|z_1)B(m^0-6|z_3)\cdots \nonumber \\ 
&&\left[ \cdots A(m^0-2l'|z_2)+\cdots \tilde {D}(m^0-2l'|z_2)
\right] |0>_{m_0}^m+\cdots .
\end{eqnarray}
\noindent The vector $|0>_{m_0}^m $ is an eigenvector of $A(m|z)$ and 
$D(m|z)$. The eigenvalues are given by (85,86). Thus we have
\begin{eqnarray}
A(m|z)|0>_{m_0}^m&=&\lambda _0(z)|0>_{m_0}^m,\nonumber \\
\tilde {D}(m|z)|0>_{m_0}^m&=&\lambda _1(z)|0>_{m_0}^m.
\end{eqnarray}
\noindent Therefore, noticing $m^0-2l'=m$, we can write $t(z)\Phi $ as
\begin{eqnarray}
t(z)&=&\left[ (\mu a'a''\cdots a^{(l')}_0)_0\lambda _0(z)
+(\mu a'\cdots a^{(l')})_1\lambda _1(z)\right] \nonumber \\
&&\times B(m^0-2|z_1)\cdots B(m|z_{l'})|0>_{m_0}^m \nonumber \\
&&+\left[ (\mu b'a_1''\cdots a_1^{(l')})_0\lambda _0(z_1)
\right. \nonumber \\
&&\left. +(\mu b'a_1''\cdots a_1^{(l')})_1\lambda _1(z_1)\right]
\nonumber \\
&&\times B(m^0-2|z)B(m^0-4|z_2)\cdots B(m|z_{l'})|0>^m_{m_0}\nonumber \\
&&+\left[ (\cdots )_0\lambda _0(z_2)+(\cdots )_1
\lambda _1(z_2)\right] B(m^0-2|z)B(m^0-4|z_1) \nonumber \\
&&\times B(m^0-6|z_3)\cdots |0>_{m_0}^m+\cdots 
\end{eqnarray}

\noindent Because of (102,103) and (104) we see that $t(z)\Phi $ must
be a linear combination of
\begin{eqnarray}
B(m^0-2|z_1)\cdots B(m|z_{l'})|0>_{m_0}^m &\equiv &\Psi _0
=\Phi \nonumber \\
B(m^0-2|z)B(m^0-4|z_2)\cdots |0>_{m_0}^m &\equiv &\Psi _1
\nonumber \\
\vdots \nonumber \\
B(m^0-2|z)B(m^0-4|z_1)\cdots 
B(\cdots |z_{i-1})B(\cdots |z_{i+1})\cdots 
|0>_{m_0}^m&\equiv &\Psi _i\nonumber \\
\vdots \nonumber \\
B(m^0-2|z)B(m^0-4|z_1)\cdots B(m|z_{l'-1})|0>_{m_0}^m
&\equiv &\Psi _{l'}
\end{eqnarray}
\noindent This is because we can always change the order of $B$'s
such that $z_i$'s inside $B$'s are arranged according to the order 
of $i$. So we have
\begin{eqnarray}
t(z)\Phi &\equiv &C_0^1\Psi _0+C_1^1\Psi _1+C_2^1\Psi _2
+\cdots +C_{l'}^1\Psi _{l'}
\end{eqnarray}
\noindent The problem now is that although the forms of $C_0^1$ and
$C_1^1$ are simple and clear, $C_i^1$ for $i\geq 2$ are
represented by a complicated summation. However, using the fact that
$\Phi $ is a symmetric function of $z_i'$s, we can greatly simplify
the calculation. Let us exchange $z_i$ and $z_1$ in $\Phi $. This does
not change $\Phi $. Then we can use the above standard procedure to have
\begin{eqnarray}
t(z)\Phi =C_0^i\Psi +C_1^i\Psi _1
+\cdots +C_i^i\Psi _i+C_{i+1}^i\Psi _{i+1}+\cdots 
\end{eqnarray}
\noindent where $C_0^i$ and $C_i^i$ can be obtained by exchanging
$z_i$ and $z_1$ in $C_0^1$ and $C_1^1$. Assume $\Psi _0,\Psi _1,
\cdots ,\Psi _{l'}$ are linearly independent vectors. Then each 
coefficient for the linear decomposition of $t(z)\Phi $ by 
$\left\{ \Psi _i\right\} $ is unique. Thus we have 
$C_i^1=C_i^i$. Put all $C_i^1=0$ for $i\not =0$ in (112) we have  
\begin{eqnarray}
t(z)\Phi =C_0^1\Phi \equiv \tau (z)\Phi .
\end{eqnarray}
\noindent It is, $\Phi $ is an eigenstate of the transfer matrix $t(z)$ 
with eigenvalue
\begin{eqnarray}
\tau (z)=\mu (z)a'a''\cdots a^{(l')}\lambda (z)
\end{eqnarray}
\noindent The spectrum parameters $z_i$ are determined by
the $l'$ conditions $C_i^1=0, i=1,\cdots,l'$. The first condition is
\begin{eqnarray}
C_1^1=\mu (z)b'a_1''\cdots a_1^{(l')}\lambda (z_1)=0.
\end{eqnarray}
\noindent Other $l'-1$ conditions can be obtained by 
exchanging $z_1$ and $z_i$ in (116). These are the Bethe ansatz equations.
Using the explicit form of (101-103) (i.e. Eqs.(56), (100), (57) and
(59)),
we can prove
that these equations are actually independent
of the spectrum parameter $z$. This implies that $\Phi $ is an 
eigenstate of all transfer matrices with arbitrary spectrum.

To end this section, we present here some results. The left boundary
matrix is diagonal, we can show that its 
diagonal elements can be explicitly written as:
\begin{eqnarray}
\tilde {K}(m^0|z)_0^0&=&h((a^0-1)w)h(\tilde {\xi }-z-w)
h((a^0+1)w+\tilde {\xi }+z)F(z),
\nonumber \\
\tilde {K}(m^0|z)_1^1&=&h((a^0+1)w)h(z+\tilde {\xi }+w)
h((a^0-1)w+\tilde {\xi }-z)F(z),
\nonumber
\end{eqnarray}

\noindent where $\tilde {\xi }$ is the left boundary free parameter. We notice that 
this solution is identified with the solution given in ref.[44]. While 
$F(z)$ is a function of $z$ depending on the scale of the left boundary matrix,
which is not essential. So we get
\begin{eqnarray}
\mu _0(z)&=&h(2z+2w)h(\tilde {\xi }-z)h(z+\tilde {\xi} +a^0w)F(z),\nonumber \\
\mu _1(z)&=h(w)&h(z+\tilde {\xi }+w)h(\tilde {\xi }-z+(a^0-1)w)F(z).
\nonumber
\end{eqnarray}
\noindent The eigenvalue of the transfer matrix of eight vertex model
with open boundary conditions is
\begin{eqnarray}
\tau (z)&=&\mu _0(z)\lambda _0(z)\prod _{i=1}^{l'}
\frac {h(z_i+z)h(z_i-z+w)}{h(z_i+z+w)h(z_i-z)}\nonumber \\
&&+\mu _1(z)\lambda _1(z)\prod _{i=1}^{l'}
\frac {h(z_i-z-w)h(z_i+z+2w)}{h(z_i+z+w)h(z_i-z)}.
\end{eqnarray}
\noindent Here $\left\{ z_i\right\} $ should satisfy 
the Bethe ansatz equations:
\begin{eqnarray}
&&\frac {\lambda _0(z_i)}{\lambda _1(z_i)}
\prod _{j=1,j\not= i}^{l'}
\frac {h(z_i+z_j)h(z_i-z_j+w)}{h(z_i-z_j-w)h(z_i+z_j+2w)}
\nonumber \\
&=&\frac {h(w)h(\tilde {\xi }-z_i+w(a^0-1))h(\tilde {\xi }+z_i+w)}
{h(2z_i)h(\tilde {\xi }+z_i+wa^0)h(\tilde {\xi }-z_i)}
\end{eqnarray}
\noindent for $i=1,\cdots ,l'$.
From (85,86) and the definitions of $A, \tilde {D}$, we have
\begin{eqnarray}
\lambda _0(z)&=&K(m_0|z)_0^0\prod _{i=1}^l\left[ 
\frac {h(z+u_i+w)h(z-u_i+w)}{[h(w)]^2}\right],\nonumber \\
\lambda _1(z)&=&\left[ K(m_0|z)_1^1-K(m_0|z)_0^0
\frac {h(2z+w(a_0+1))h(w)}{h(w(a_0+1))h(2z+w)}\right] 
\frac {h(2z+w)h(w(a_0+1))}{h(w)h(wa_0)} \nonumber \\
&&\times \prod _{i=1}^l\left[ \frac {h(z+u_i)h(z-u_i)}   
{[h(w)]^2}\right] ,
\nonumber
\end{eqnarray}

\noindent where $a^0\equiv m^0+\gamma =a+2l'$, $a_0\equiv m_0+\gamma 
\equiv a+l$, $a\equiv m+\gamma $.

\section{Discussions}
We can obtain our trigonometric limit as the following. The intertwiner
(or three spin operator) defined in (27) is 
\begin{eqnarray}
\phi _{m,\mu }^k(z)=\theta \left[ \begin{array}{c}
{1\over 2}-{k\over 2} \\ {1\over 2} \end{array}\right]
(z+(-1)^{\mu }w(m+\gamma )+w\beta ,2\tau )
\end{eqnarray}
\noindent Define $\gamma '=\gamma +{\tau \over 2w}, 
\beta '=\beta +{\tau \over 2w}$ and note $\mu =0,1$. Equation (27) reads
\begin{eqnarray}
\phi _{m,\mu }^k(z)&=&\theta \left[ \begin{array}{c}
\frac {1}{2}-\frac {k}{2} \\ \frac {1}{2} \end{array} \right]
(z+(-1)^{\mu }w(m+\gamma ')
+w\beta '+(\mu -1)\tau ,2\tau )\nonumber \\
&=&\xi (\mu )\theta \left[
\begin{array}{c} 
\frac {\mu -k}{2}\\
\frac {1}{2}\end{array} \right] 
(z+(-1)^{\mu }w(m+\gamma ')+w\beta ',2\tau ),
\end{eqnarray}
\noindent where 
\begin{eqnarray}
\xi (\mu )=e^{-2\pi i(\frac {\mu -1}{2})
(\frac {\mu -1}{2}\tau +z+(-1)^{\mu }w(m+\gamma ')+w\beta '+\frac {1}{2})}
\end{eqnarray}
\noindent is independent of $k$. We then rescale $\phi $ to 
$\phi '=\xi ^{-1}\phi $. At the same time, we must perform a gauge
transformation which change $W$ ot $W'$ to ensure the face-vertex
correspondence. When $\tau \rightarrow i\infty ,
{\phi '}_{m,\mu }^k(z)\rightarrow \delta _{\mu k}$. $W'$ goes to a
trigonometric R matrix, which is different with the R matrix in ref.[16]
only by a constant factor. The boundary condition $K(m|z)_1^0=0$,
$\tilde {K}(m|z)_1^0=\tilde {K}(m|z)_0^1=0$ (if we add 
$K(m|z)_0^1=0$) also approach that of ref.[16]. Thus we can show that 
the trigonometric limit of our model is that of ref.[16]. It is a 
six vertex model with integrable reflection boundaries. In such model,
the number of $B$'s ($l'$) in the Bethe ansatz state $\Phi $ is
arbitrary. It is reasonable that the eigenstate of transfer matrix with
maximum absolute value of eigenvalue (ETMM) is a Bethe ansatz state. Each
such six vertex model can be attained as a limit of a sequence of eight
vertex model with reflection boundaries. In this limit procedure, the
Bethe ansatz equations, vaccum states and operators $A,B,C,D$ are
all approaching that of six vertex model. Thus we can reasonably assume
in this sequence of eight vertex models there is a sequence of Bethe 
ansatz states which approach the ETMM of the six vertex model. Thus
it is quite possible that these Bethe ansatz states are ETMM of the
eight vertex models, especially when they are very close to the limit
six vertex model since the eigenvalues of the transfer matrix $t(z)$ are
discrete[7]. The fact whether a Bethe ansatz state is with the
maximum absolute value of eigenvalue, should not depend on continuous
parameters if there is no phase transition in the procedure, also 
since eigenvalues are discrete. For the above six vertex
model, when $l$ is given (the column number is given), the true
discrete varible is $l'$. The above sequence of Bethe ansatz states should
have same $l'$. Thus, it is reasonable that in our approach, when the
left and right boundary condition determine a proper $l'$, the
Bethe ansatz state has a eigenvalue of maximum absolute value, which is
the most important state in thermodynamics.

We know that Johnson, Krinsky and McCoy calculated the energy of
excitations of the $XYZ$ model after Baxter obtains the Bethe ansatz
of eight vertex model with periodic boundary conditions[48]. Now, using
the results presented in this paper, we may also calculate the
energy of the excitations of the $XYZ$ model with boundaries. By 
analyzing the eigenvalues of the transfer matrix, one may also 
get the boundary free energy, thermodynamic limit and finite size 
corrections. Other physical phenomena are also worth of
studying suh as surface critical exponents and scaling, the central
charges in conformal field theory etc.. It is well known that eight
vertex model is equivalent to SOS model, if one impose some restrictions
on SOS model, we can obtain the restricted SOS model (ABF model). 
So, if we impose some restrictions on the Bethe ansatz of eight vertex model
with boundaries, we should find the Bethe ansatz for the ABF model with 
boundary conditions. All of these work worth be studied in the future.

\vskip 2truecm
\noindent {\large \bf Acknowlegements:} We wish to thank 
Professors P.A.Pearce and Y.K.Zhou for their very helpful discussions
and their valuable preprints. One of us, Hou would like to thank
Prof.Pearce for his hospitality in Melborn.

\newpage

\section*{Appendix A $~~~$ Quadratic relation of the components of the 
face type $k(z)$}
The left hand side of equation (24) is 
$R_{12}(z_1-z_2)k_1(z_1)R_{21}(z_1+z_2)k_2(z_2)$. We multiply it from 
left by $\tilde {\phi }_{m+\hat {\mu }_0,\mu _0}^{(1)}(z_1)
\tilde {\phi }_{m+\hat {\mu }_0+\hat {\nu }_0,\nu _0}^{(2)}(z_2)$
and multiply it from right by $\phi _{m+\hat {\mu }_3,\mu _3}^{(1)}
(-z_1)\phi _{m+\hat {\mu }_3+\hat {\nu }_3,\nu _3}^{(2)}(-z_2)$ obtaining
\begin{eqnarray}
LHS=\tilde {\phi }_{m+\hat {\mu }_0,\mu _0}^{(1)}(z_1)
\tilde {\phi }_{m+\hat {\mu }_0+\hat {\nu }_0,\nu _0}^{(2)}(z_2)
R_{12}(z_1-z_2)\cdots .
\end{eqnarray}
\noindent Using (34) we can eliminate $R_{12}$ to have 
\begin{eqnarray}
LHS&=&W(m|z_1-z_2)_{\mu _0\nu _0}^{\mu _1\nu _1}
\tilde {\phi }_{m+\hat {\mu }_1+\hat {\nu }_1,\mu _1}^{(1)}(z_1)
\tilde {\phi }_{m+\hat {\nu }_1,\nu _1}^{(2)}(z_2)k_1(z_1)\nonumber \\
&&R_{21}(z_1+z_2)k_2(z_2)\phi _{m+\hat {\mu }_3,\mu _3}^{(1)}(-z_1)
\phi _{m+\hat {\mu }_3+\hat {\nu }_3,\nu _3}^{(2)}(-z_2).
\end{eqnarray}
\noindent Move $\tilde {\phi }^{(2)}$ over $k_1(z_1)$ and move 
$\phi ^{(1)}$ over $k_2(z_2)$, so that they are at the left and right
side of $R_{21}$ respectively. LHS becomes
\begin{eqnarray}
LHS&=&\cdots \tilde {\phi }_{m+\hat {\nu }_1,\nu _1}^{(2)}(z_2)
R_{21}(z_1+z_2)\phi _{m+\hat {\mu }_3,\mu _3}^{(1)}(-z_1)\cdots \nonumber \\
&=&\cdots W(m|z_1+z_2)^{\nu _2\mu _3}_{\nu _1\mu _2}
\phi _{m+\hat {\mu }_2+\hat {\nu }_1,\mu _2}^{(1)}(-z_1)
\tilde {\phi }_{m+\hat {\nu }_1+\hat {\mu }_2,\nu _2}^{(2)}(z_2)\cdots .
\nonumber \\
\end{eqnarray}
\noindent Since the Boltzmann weight $W$ is non zero only if 
$\hat {\nu }_2+\hat {\mu }_3=\hat {\nu }_1+\hat {\mu }_2$, we have
\begin{eqnarray}
LHS=\cdots \phi _{m+\hat {\mu }_2+\hat {\nu }_1,\mu _2}^{(1)}(-z_1)
W(m|z_1+z_2)_{\nu _1\mu _2}^{\nu _2\mu _3}
\tilde {\phi }_{m+\hat {\mu }_3+\hat {\nu }_2,\nu _2}^{(2)}(z_2)\cdots .
\end{eqnarray}
\noindent Now $k_1(z_1)$ and $k_2(z_2)$ all have their "own" 
$\tilde {\phi }$ and $\phi $ at left and right side. By definition one
conclude
\begin{eqnarray}
LHS&=&k(m+\hat {\mu }_2+\hat {\nu }_1|z_1)_{\mu _1}^{\mu _2}
k(m+\hat {\mu }_3+\hat {\nu }_3|z_2)_{\nu _2}^{\nu _3}\nonumber \\
&&\times W(m|z_1-z_2)_{\mu _0\nu _0}^{\mu _1\nu _1}
W(m|z_1+z_2)_{\nu _1\mu _2}^{\nu _2\mu _3}.
\end{eqnarray}
\noindent The derivation of RHS is  similar.

\section*{Appendix B $~~~$ Left and Right inverse matrices of $k(z)$}
When $n=2$, we have 
\begin{eqnarray}
R_{12}(z)=R_{21}(z)=R_{12}(z)^{t_1t_2}.
\end{eqnarray}
\noindent When $z=-w$, we have
\begin{eqnarray}
P_-(12)R_{12}(-w)=R_{12}(-w)P_-(12)=R_{12}(-w),
\end{eqnarray}
\noindent where $P_-(12)$ is the anti-symmetric operator of space 
$V_1\otimes V_2$ satisfying $P_-(12)^2=P_-(12)$. On the other hand the
R matrix have the property
\begin{eqnarray}
I_{\alpha }^{(i)}I_{\alpha }^{(j)}R_{ij}(z)[I_{\alpha }^{(i)}]^{-1}
[I_{\alpha }^{(j)}]^{-1}=R_{ij}(z).
\end{eqnarray}
\noindent Thus
\begin{eqnarray}
&&I_{\alpha }^{(3)}P_-(12)R_{32}(z_2)R_{31}(z_1)P_-(12)
[I_{\alpha }^{(3)}]^{-1}\nonumber \\
&=&P_-(12)I_{\alpha }^{(3)}R_{32}(z_2)[I_{\alpha }^{(3)}]^{-1}
I_{\alpha }^{(3)}R_{31}(z_1)[I_{\alpha }^{(3)}]^{-1}P_-(12)\nonumber \\
&=&P_-(12)[I_{\alpha }^{(2)}]^{-1}R_{32}(z_2)I_{\alpha }^{(2)}
[I_{\alpha }^{(1)}]^{-1}R_{31}(z_1)I_{\alpha }^{(1)}P_-(12)\nonumber \\
&&=P_-(12)[I_{\alpha }^{(2)}]^{-1}[I_{\alpha }^{(1)}]^{-1}
R_{32}(z_2)R_{31}(z_1)I_{\alpha }^{(2)}I_{\alpha }^{(1)}P_-(12).
\end{eqnarray}
\noindent It is not difficult to show
\begin{eqnarray}
I_{\alpha }^{(1)}I_{\alpha }^{(2)}P_-(12)=(-1)^{\alpha _1+\alpha _2}
P_-(12)=P_-(12)[I_{\alpha }^{(1)}]^{-1}[I_{\alpha }^{(2)}]^{-1}.
\end{eqnarray}
\noindent Thus
\begin{eqnarray}
&&I_{\alpha }^{(3)}P_-(12)R_{32}(z_2)R_{31}(z_1)P_-(12)
[I_{\alpha }^{(3)}]^{-1}\nonumber \\
&=&P_-(12)R_{32}(z_2)R_{31}(z_1)P_-(12)\equiv U.
\end{eqnarray}
\noindent The operator $U$ acting on $V_1\otimes V_2\otimes V_3$ is
invariant under the similar transformation by 
$I_{\alpha }^{(3)}$, which implies that $U$ is equivalent to a unit 
operator on the $V_3$. From (127) we see that this is also true for
$U'=P_-(12)R_{23}(z_2)R_{13}(z_1)P_-(12)$. We then consider the RE (14) 
for the case $z_1=z-w, z_2=z$, and have
\begin{eqnarray}
G&=&R_{12}(-w)k_1(z-w)R_{21}(2z-w)k_2(z)\nonumber \\
&=&R_{12}(-w)T_1(z-w)K_1(z-w)S_1(z-w)R_{21}(2z-w)T_2(z)
K_2(z)S_2(z)\nonumber 
\end{eqnarray}
\noindent Due to (21), 

\begin{eqnarray}
G&=&R_{12}(-w)T_1(z-w)K_1(z-w)T_2(z)R_{21}(2z-w)S_1(z-w)
K_2(z)S_2(z)\nonumber \\
&=&R_{12}(-w)T_1(z-w)T_2(z)K_1(z-w)R_{21}(2z-w)K_2(z)S_1(z-w)S_2(z).
\nonumber \\
\end{eqnarray}
\noindent Rewrite $R_{12}(-w)$ as $P_-(12)R_{12}(-w)$, and move 
$R_{12}(-w)$ towards the right. Each time when it goes over a pair
of $R_{1i}R_{2i}$ in $T_1T_2$ by YBE (9), we rewrite $R_{12}(-w)$ as 
$P_-^2(12)R_{12}(-w)$ and leave $P_-^2$. Then we obtain
\begin{eqnarray}
G&=&\left[ P_-(12)R_{23}R_{13}P_-(12)R_{24}R_{14}P_-(12)
\cdots P_-(12)R_{2l}R_{1l}P_-(12)\right]\nonumber \\
&&R_{12}(-w)K_1(z-w)R_{21}(2z-w)K_2(z)S_1(z-w)S_2(z)\nonumber \\
&\equiv &[M]\times \cdots .
\end{eqnarray}
\noindent Move $R_{12}(-w)$ to the left side of $S_1$ by RE. One has
\begin{eqnarray}
G=[M]K_2(z)R_{12}(2z-w)K_1(z-w)R_{21}(-w)S_1(z-w)S_2(z).
\end{eqnarray}
\noindent Similarly, we move $R_{21}(-w)$ step by step to the right
side of $S_1S_2$ obtaining
\begin{eqnarray}
G&=&[M]\cdots [N]R_{21}(-w)\nonumber \\
\left[ N\right]&=&[P_-(12)R_{l2}R_{l1}
P_-(12)P_-(12)R_{l-1,2}R_{l-1,1}P_-(12)\cdots
\nonumber \\
&&P_-(12)R_{32}R_{31}P_-(12)].
\end{eqnarray}
\noindent From the property of $U$ and $U'$, we see that $G$ is
proportional to an identity operator in the "quantum" space 
$V'=V_3\otimes \cdots \otimes V_l$. Using the similar derivation as
in Appendix A, we multiply 
$\tilde {\phi }^{(1)}\tilde {\phi }^{(2)}$ from left and multiply 
$\phi ^{(1)}\phi ^{(2)}$ from right of $G$, and conclude that when
$z_1=z-w, z_2=z$, both LHS and RHS of (40) are proportional to 
identity operator in the quantum space $V'$. Properly choosing indices
and noticing $W(m|-w)_{\mu \nu }^{\mu '\nu '}=\cdots
\delta _{\mu \bar {\nu }}\delta _{\mu '\bar {\nu }'}$, we have (49).

From the above derivation, we see also that $G$ is anti-symmetric to
the classical (auxiliary) indices of space $V_1\otimes V_2$ and
is independent of $m$. Thus
\begin{eqnarray}
\rho '(m,\mu |z)&=&
\tilde {\phi }_{m+\hat {\bar {\mu }},\bar {\mu }}^i(z-w)
\tilde {\phi }^j_{m+\hat {\bar {\mu }}+\hat {\mu },\mu }(z)
G_{ij}^{i'j'}\nonumber \\
&&\times \phi _{m+\hat {1},1}^{i'}(-z+w)
\phi _{m+\hat {1}+\hat {0},0}^{j'}(-z)\nonumber \\
&=&\left\{ \tilde {\phi }^1_{\cdots }(z-w)\tilde {\phi }^0_{\cdots }(z)
-\tilde {\phi }^0_{\cdots }(z-w)\tilde {\phi }^1_{\cdots }(z)\right\} 
\nonumber \\
&&\left\{ \phi ^1_{\cdots }(-z+w)\phi ^0_{\cdots }(-z)
-\phi ^0_{\cdots }(-z+w)\phi ^1_{\cdots }(-z)\right\} 
G_{10}^{10}.
\end{eqnarray}
\noindent Similarly, $\rho (m,\nu |z)$ can also be expressed as
$G_{10}^{10}$ multiplied by a factor 
which depends only on $\phi ,\tilde {\phi }$. 
Therefore the ratio $\frac {\rho (m+2|z)_0}{\rho '(m|z)_0}$ is
independent of $G_{10}^{10}$. It is completely determined by 
$\phi ,\tilde {\phi }$. Direct calculation gives (53).

\section*{Appendix C $~~~$ Derivation of the boundary condition}
From definition we have
\begin{eqnarray}
\phi _{m,\mu }^k(z)&=&\theta \left[ \begin{array}{c}
{1\over 2}-{k\over 2}\\ {1\over 2}\end{array}
\right] (z+(-1)^{\mu }wa +w\beta ,2\tau )\nonumber \\
&\equiv &\theta ^{(k)}(z+w((-1)^{\mu }a+\beta ))\nonumber \\
&\equiv &\theta ^{(k)}(z+\chi ),
\end{eqnarray}
\noindent  where $a\equiv m+\gamma $. When $\alpha _1,\alpha _2$ are 
integers, by the expression of $\theta $ function, we have
\begin{eqnarray}
&&\theta ^{(k)}(z+\chi +\alpha _1\tau +\alpha _2)\nonumber \\
&=&e^{-2\pi i(\frac {\alpha _1}{2})
(\frac {\alpha _1}{2}\tau +z+\chi +{1\over 2})+2\pi i({1\over 2}-
{k\over 2})\alpha _2}\theta ^{(k-\alpha _1)}(z+\chi ),
\end{eqnarray}
\noindent giving
\begin{eqnarray}
h^{\alpha _1}g^{\alpha _2}\phi _{m,\mu }(z)=
(-1)^{\alpha _2}e^{2\pi i({{\alpha _1}\over 2})
(\frac {\alpha _1}{2}\tau +z+\chi +{1\over 2}+\alpha _2)} 
\phi _{m,\mu }(z+\alpha _1\tau +\alpha _2).
\end{eqnarray}
\noindent On the other hand, from the property of zeros of doublely 
quasi-periodic holomorphic funcion, one can show[49]
\begin{eqnarray}
Det\left[ \begin{array}{cc}
\theta ^{(0)}(z_1) &\theta ^{(0)}(z_2)\\
\theta ^{(1)}(z_1) &\theta ^{(1)}(z_2) \end{array}\right] 
=C\times h(\frac {z_1+z_2-1}{2})h(\frac {z_1-z_2}{2}),
\end{eqnarray}
\noindent where $C$ is independent of $z_1$ and $z_2$. If we write
\begin{eqnarray}
A\equiv \left[ \begin{array}{cc}
\phi _{m+\hat {0},0}^{0}(z) &\phi _{m+\hat {1},1}^0(z)\\
\phi _{m+\hat {0},0}^1(z) &\phi _{m+\hat {1},1}^1(z)
\end{array} \right]
\end{eqnarray}

\noindent then $\tilde {\phi }_{m+\hat {\mu },\mu }^k(z)$ are elements of
its inverse matrix. Thus for any column vector $\Psi $, the quantity
$B\equiv \sum _k\tilde {\phi }_{m+\hat {1},1}^k(z)\Psi ^k$ can 
be written as 
\begin{eqnarray}
B=\frac {1}{Det{\rm A}}Det \left[
\begin{array}{cc}
\phi _{m+\hat {0},0}(z) &{\Psi ^0}\\
\phi _{m+\hat {0},0}(z) &{\Psi ^1}\end{array}\right] .
\end{eqnarray}
\noindent Combining (140) and (143) gives
\begin{eqnarray}
B'_{\alpha }&\equiv &\tilde {\phi }_{m-2,1}(z)h^{\alpha _1}
g^{\alpha _2}\phi _{m,0}(-z)\nonumber \\
&=&\left( \frac {1}{Det{\rm A'}}\right) Det\left[ \begin{array}{cc} 
\theta ^{(0)}(z+wa+w\beta ) &
\theta ^{(0)}(-z+wa+w\beta +\alpha _1\tau +\alpha _2)\\
\theta ^{(1)}(z+wa+w\beta ) &
\theta ^{(1)}(-z+wa+w\beta +\alpha _1\tau +\alpha _2)\end{array}
\right]  \nonumber \\
&&\times e^{2\pi i({\alpha _1 \over 2})({\frac {\alpha _1\tau }{2}}
-z+wa+w\beta +{1\over 2}+\alpha _2)}\times (-1)^{\alpha _2}.
\end{eqnarray}
\noindent Two determines can be obtained from (141), thus
\begin{eqnarray}
B'_{\alpha }=
-\frac {\left[ 
\sigma _{\alpha }(wa+w\beta -\frac {1}{2})\sigma _{\alpha }(-z)
(-1)^{\alpha _2}\right] }
{[h(z+w\beta +w-\frac {1}{2})h(wa-w)]}
\end{eqnarray}
\noindent Substituting the definition of $K(m|z)_1^0$ and the expression
of $K(z)$(15,63), we have
\begin{eqnarray}
K(m|z)_1^0=(-1)\frac {\sum _{\alpha }C_{\alpha }(-1)^{\alpha _2}
\sigma _{\alpha }(wa+w\beta -\frac {1}{2})}
{h(z+w\beta +w-\frac {1}{2})h(wa-w)}.
\end{eqnarray}
\noindent Other elements of $K(m|z)$ can be similarly obtained. They are.

\noindent $K(m|z)_0^1=a\rightarrow -a$ in RHS of the above equation.
\begin{eqnarray}
K(m|z)_0^0&=&\frac {1}{h(z+w\beta +w-{1\over 2})h(wa-w)}\nonumber \\
&&\times \sum _{\alpha }C_{\alpha }(-1)^{\alpha _2}
\frac {\sigma _{\alpha }(w\beta +w-{1\over 2})
\sigma _{\alpha }(-z+wa-w)}{\sigma _{\alpha }(-z)} \nonumber \\
\end{eqnarray}
\noindent $K(m|z)_1^1=a\rightarrow -a$ in RHS of equation (147).
\begin{eqnarray}
\tilde {K}(m|z)_1^0=\frac {1}{h(-z+w\beta -{1\over 2})h(wa)}
\sum _{\alpha }\tilde {C}_{\alpha }(-1)^{\alpha _2}
\sigma _{\alpha }(-wa+w+w\beta -{1\over 2}) \nonumber \\
\end{eqnarray}
\noindent $\tilde {K}(m|z)_0^1=a\rightarrow -a$ in RHS of 
equation (148).
\begin{eqnarray}
\tilde {K}(m|z)_0^0=\frac {1}{h(-z+w\beta -{1\over 2})h(wa)}
\sum _{\alpha }\tilde {C}_{\alpha }(-1)^{\alpha _2}
\frac {\sigma _{\alpha }(w\beta -{1\over 2})
\sigma _{\alpha }(z+wa)}{\sigma _{\alpha }(z+w)} \nonumber \\
\end{eqnarray}
\noindent $\tilde {K}(m|z)_1^1=a\rightarrow -a $ in RHS of equation (149).

\end{document}